# A discrete-module-finite-element hydroelasticity method in analyzing dynamic response of floating flexible structures

Yongqiang Chen[1], Xiantao Zhang[1,2*], Lei Liu[1,2], Xinliang Tian[1,2], Xin Li[1,2], Zhengshun Cheng[1,2]

1 State Key Laboratory of Ocean Engineering, Shanghai Jiao Tong University, Shanghai 200240, China

2 SJTU Sanya Yazhou Bay Institute of Deepsea Technology, Sanya 572024, China

**Abstract**: As an alternative to mode-superposition or direct method in hydroelastic analysis, a discrete-module-finite-element (DMFE) based hydroelasticity method with features of avoiding pre-determination of structure modal shapes and better computational efficiency has been proposed and well developed. The DMFE solves the hydroelastic problem as follows. Firstly, a freely floating flexible structure is discretized into several macro-submodules in two horizontal directions to perform a multi-rigid-body hydrodynamic analysis. Each macro-submodule is then abstracted to a lumped mass at the center of gravity that bears the external forces including inertia force, hydrodynamic force and hydrostatic force (with displacements as unknown variables) acting on the macro-submodule. Apart from external forces, all lumped masses are also subjected to structural forces that reflect the structural deformation features of the original flexible structure. The key to calculating the structural forces is derivation of the equivalent overall structural stiffness matrix with respect to the displacements of all lumped masses, which is tackled following the finite element procedure. More specifically, each macro-submodule is discretized into a number of micro-elements to derive the corresponding structural stiffness matrix, which is manipulated to a new one including only the nodes at the position of the lumped masses and surrounding boundaries by using the substructure approach, and subsequently the target overall stiffness matrix is obtained by combining together all macro-submodules. Finally, based on equivalence between external and structural forces indicated from the d'Alembert principle, the DMFE method establishes the hydroelastic equation on all lumped masses with their displacements as unknown variables. Solving the equation gives the displacement response of all lumped masses. Displacement and structural force responses are first calculated on the interfaces of every two adjacent macro-submodules, after which at any given position of the flexible structure, the recovery of displacement is based on the structural stiffness matrix of the corresponding macro-submodule and the recovery of structural force uses the spline interpolation scheme. The hydroelasticity of a narrow and a square pontoon-type VLFS is investigated. Satisfactory agreement is achieved with numerical results from other scholars. At last, a least square method to recover bending moment distribution of flexible floating structures with complicated shape is presented, including some unsolved problems.

# 1 Introduction

Very large floating structure (VLFS), which can be used as floating pier, floating airport, floating solar or even floating city, has triggered extensive research [1]. Hydroelasticity of VLFSs under wave loads is a typical Fluid-Structure-Interaction (FSI) problem [2]. The hydroelasticity theory is adopted to calculate the dynamic response of a floating flexible structure, whose core idea is to solve the hydrodynamic equation that consists of inertia force, hydrostatic and hydrodynamic loadings, and force due to elastic deformation [3]. The modal analysis method and the direct method occupy a dominant position in studying the hydroelastic response of floating flexible structures. The former solves the problem by superposing the dynamic response of necessary oscillation modes of a floating flexible structure, where a pre-analysis is needed to determine the optimal combination of modes [4-6]. The latter is gradually adopted with increasingly powerful computing performance, although it is still time-consuming to analyze a single case [7-8].

In recent years, an efficient and convenient hydroelasticity method, i.e., the discrete-module-beam (DMB) method, has caught researchers' attention since the work of Lu et al. [9], where a continuous floating flexible structure was studied. The DMB method first discretizes a floating flexible structure into several rigid macro-submodules to perform a multi-rigid-body hydrodynamic analysis, with which the external loading on the macro-submodules (with displacements as unknown variables) can be derived and the macro-submodules are abstracted as lumped masses. Then, an equivalent Euler-Bernoulli beam that carries the flexible structure's properties, including elastic modulus, poison's ratio, etc., is introduced to connect every adjacent lumped mass, accounting for the structural deformation effects. Finally, the hydroelasticity equation is established and solved to obtain the dynamic response of the floating flexible structure. Progress has been made on the DMB hydroelasticity method by many researchers in different aspects. Sun et al. [10] explored a hinged VLFS and proposed a third-order interpolation scheme to calculate the bending moment distribution in the framework of DMB method. Zhang and Lu [11] proposed an approach to obtain the stiffness matrix of VLFSs with complex geometric features. Further, Zhang et al. [12] extended the DMB method into time domain and calculated the displacement response of a VLFS under weight-drop loads and moving point loads.

Many researchers have applied the DMB method to investigate different engineering problems. Wei et al. [13] analyzed a VLFS's hydroelastic behavior in inhomogeneous sea conditions. Zhang et al. [14] and Lu et al. [15] investigated the dynamic response and power capture performance of a floating flexible structure with a wave energy conversion unit. Jin et al. [16] implemented the DMB method to solve a floater-connector-mooring coupled system. Bakti et al. [17] considered the forward speed effect in the hydroelasticity problem of a vessel. Zhang et al. [18] explored the hydroelasticity of a VLFS where a certain number of wind turbines are placed.

By far, the framework of the DMB method has been well established. It is worth noting that, all above-mentioned research work concerned with the DMB method focused on floating flexible structures with a relatively large length/width ratio, for the reason that the DMB method is a 'one-dimensional' method, where submodule-division is done only longitudinally (this is what we mean by stating 'one-dimensional' and a detailed explanation will be given in Section 3) and beam elements are adopted. Although it has been proven and recognized the method's accuracy in calculating hydroelastic response and adaptability in different engineering scenarios, the method itself still remains a 'one-dimensional' method, which limits its applications only in narrow structures that has a relatively large length/width ratio. Besides, even for a narrow structure case, it cannot be given by the DMB method the displacement and bending moment distribution along the width direction. However, for a floating flexible structure with a comparable length/width ratio that is common in the realm of ocean engineering, deformation and internal force response matter in both directions (length and width), whereas the DMB method cannot deal with such kind of problems. Therefore, to fill the gap, a discrete-module-finite-element (DMFE) based hydroelasticity method is developed that is capable of addressing the hydroelasticity of large floating flexible structures of arbitrary shape and size ratio.

The remainder of this paper is organized as follows. The DMFE method is first elaborated in Section 2, including the discretization strategy, derivation of the lumped-mass stiffness matrix, recovery of the displacement and internal force responses. Comparisons and relations of the DMFE method with the DMB method are given in Section 3. Validations and applications of the DMFE method are given in Section 4 on both a narrow VLFS and a square one. Besides, more hydroelasticity results on an author-defined square VLFS are given in the Appendix. At last, an unsolved problem is elaborated in Section 5 on the bending moment distribution.

## 2 The DMFE method

The DMFE method focuses on the hydroelasticity of a floating flexible structure. This problem is elaborated in Fig. 1. A pontoon type VLFS with the size of $L \times B \times D$ is subjected to an incident wave coming with an angle of $\theta$. The VLFS is uniform and has a homogeneous, isotropic and linear material. A right-handed global coordinate system $O - XYZ$ is adopted with the $Z$ axis pointing upward and the origin $O$ located at the still water level vertically and at the left lower corner of the structure horizontally. Potential flow theory is used with the fluid considered inviscid, irrotational and incompressible.

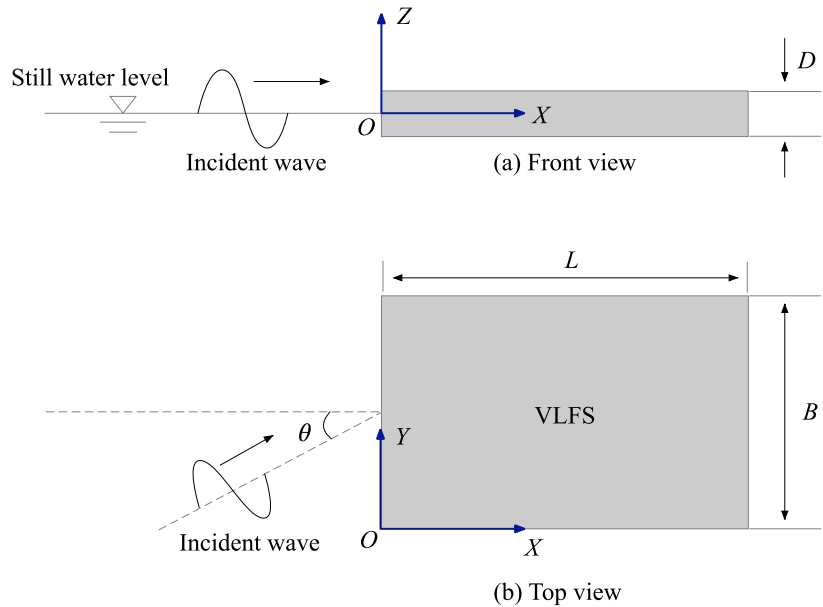

**Fig. 1 A schematic of the hydroelastic problem of a pontoon type VLFS subjected to waves.**

A brief illustration of the procedure of the DMFE method are shown in Fig. 2. Two 'discretization' strategies are adopted for the hydroelastic analysis. The first one is related to the 'macro-submodule division' of the floating flexible structure, in which the structure is divided into several submodules in both the length and width directions. Multi-rigid-body hydrodynamic analysis is conducted to obtain the hydrodynamic loadings, i.e., the wave excitation force $\mathbf{F}_E$, added mass force $\mathbf{F}_A$ and radiation damping force $\mathbf{F}_{Rd}$, which, together with the hydrostatic force $\mathbf{F}_{HS}$ and inertial force $\mathbf{F}_{In}$, comprise the total external force $\mathbf{F}_{EXT}$ (with displacements $\boldsymbol{\xi}$ as unknown variables) exerted on all the submodules. All submodules are then abstracted to be lumped masses (at centers of gravity of the submodules) that are subjected to the total external force. The second 'discretization' is the 'finite element discretization' of the floating flexible structure, the purpose of which is to derive the lumped-mass stiffness matrix $\mathbf{K}$ with respect to all lumped masses. The structural force $\mathbf{F}_{St}$ on all lumped masses can be derived based on the lumped-mass stiffness matrix and the unknown displacement variables. Finally, the hydroelastic equation of the floating flexible structure is established (at positions of all lumped masses, with displacements as unknown variables) according to the force equilibrium condition following the d'Alembert principle.

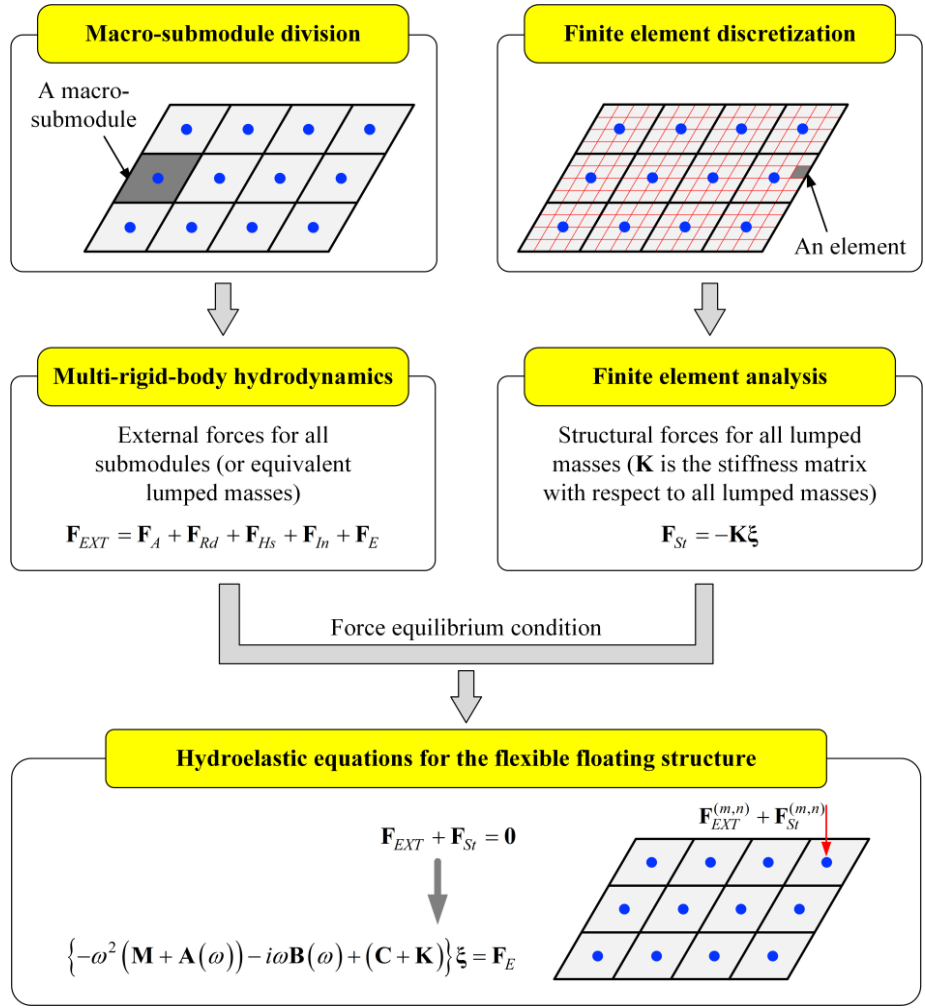

Fig. 2. The flow diagram of the DMFE method. The definition of the variables shown in the figure will be illustrated in detail in the following subsections. The floating flexible structure is shown from top view.

## 2.1 Macro-submodule division

The macro-submodule division strategy is illustrated in Fig. 3 that the floating flexible structure is divided (or discretized) equally into $M$ parts along the $OX$ axis and $N$ parts along the $OY$ axis. A total of $M \times N$ (written as $MN$ for the sake of brevity) macro-submodules are obtained. Each macro-submodule is labelled with a unique coordinate $(m, n)$ with $m$ being the $m^{th}$ column in the macro-submodule-array and $n$ the $n^{th}$ row. Since the division is done uniformly, every macro-submodule has a dimension of $L/M \times B/N \times D$. Further, the boundaries of certain macro-submodules form the 'interface', the solid line framed in dashed red box as an example. Division strategy in Fig. 3 leaves $M + 1$ interfaces parallel to the $Y$ axis and $N + 1$ interfaces parallel to the $X$ axis. Particularly, four interfaces among them are free ends, which are $X = 0$, $X = L$, $Y = 0$ and $Y = B$.

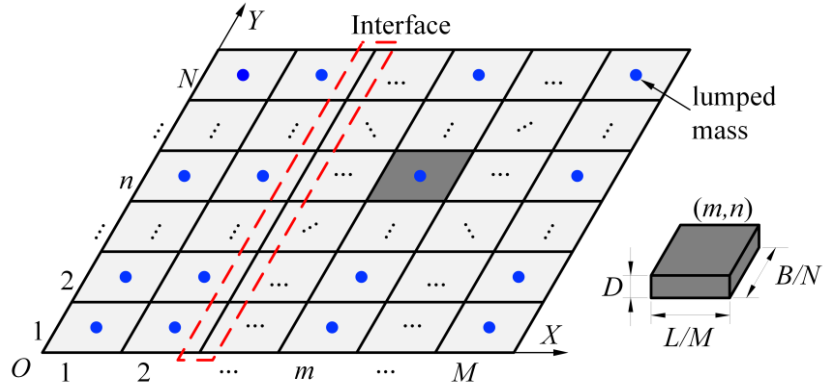

**Fig. 3. Macro-submodule division in the DMFE method. The structure is divided into $M$ and $N$ parts in the $X$ and $Y$ direction, respectively, leaving each macro-submodule being a $L/M \times B/N \times D$ cuboid. The solid line framed in dashed red box is an 'interface' composed of boundaries of certain macro-submodules.**

Each macro-submodule is then abstracted to be a lumped mass at the center of gravity that bears the external forces including inertia force, hydrodynamic force and hydrostatic force (with displacements as unknown variables) acting on the macro-submodule. Calculations on the external forces will be given in Section 2.2.

Fig. 4 gives the three right-handed coordinate systems adopted in the DMFE method, that is, the global (earth-fixed) coordinate system $O - XYZ$, the reference coordinate system $O'_{(m,n)} - x'_{(m,n)}y'_{(m,n)}z'_{(m,n)}$ and the body-fixed coordinate system $O_{(m,n)} - x_{(m,n)}y_{(m,n)}z_{(m,n)}$. Each macro-submodule $(m, n)$ has a reference coordinate system and a body-fixed one with the origins located at the corresponding center of gravity when the structure stays at its equilibrium position. When the structure is subjected to waves, the reference coordinate system stays parallel to the global one and remains fixed at its original position, while the body-fixed coordinate system moves and rotates together with the corresponding macro-submodule.

Since each macro-submodule is considered rigid, the displacement at its center of gravity, i.e., where the lumped mass is assumed to be located, $\boldsymbol{\xi}^{(m,n)}$ is used to describe the motion of macro-submodule $(m, n)$, which is expressed as

$$\boldsymbol{\xi}^{(m,n)} = \begin{bmatrix} \xi_1^{(m,n)} & \xi_2^{(m,n)} & \xi_3^{(m,n)} & \xi_4^{(m,n)} & \xi_5^{(m,n)} & \xi_6^{(m,n)} \end{bmatrix} \quad (1)$$

where $\xi_j^{(m,n)} (j = 1,2,3)$ are three translational displacements in the reference coordinate system and $\xi_j^{(m,n)}(j = 4,5,6)$ three rotational ones.

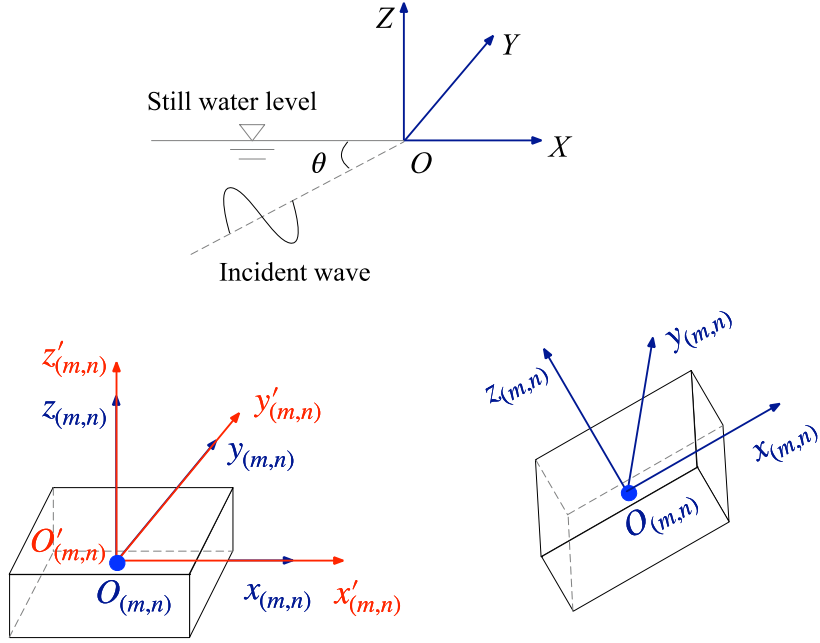

**Fig. 4.** Three coordinate systems adopted in the DMFE method, i.e., the global coordinate system $O - XYZ$, the reference coordinate system $O'_{(m,n)} - x'_{(m,n)}y'_{(m,n)}z'_{(m,n)}$ and the body-fixed coordinate system $O_{(m,n)} - x_{(m,n)}y_{(m,n)}z_{(m,n)}$.

## 2.2 Multi-rigid-body hydrodynamic analysis

Linear assumptions and harmonic excitations make it feasible to express a spatial-temporal variable $\Theta(x,y,z,t)$ as follows

$$\Theta(x,y,z,t) = \Xi(x,y,z)e^{-i\omega t} \tag{2}$$

where $\Xi(x,y,z)$ is a space-dependent complex variable, $i$ the imaginary unit, $\omega$ the circular frequency (of incident waves, as an example) and $t$ the time instant. Only the space-dependent variable is considered in the frequency domain analysis.

The velocity potential $\phi(x,y,z)$ stands under linear and ideal fluid assumptions, which is expressed in the global coordinate system as

$$\phi(x,y,z) = \phi_I + \phi_S + \sum \phi_R^{(m,n)} \tag{3}$$

where $\phi_I$ is the velocity potential of incident waves, $\phi_S$ the velocity potential of scattered waves and $\phi_R^{(m,n)}$ the velocity potential of radiated waves induced by macro-submodule $(m,n)$, which can be further written as

$$\phi_R^{(m,n)} = \sum_{j=1}^{6} \phi_{jR}^{(m,n)} = -i\omega \sum_{j=1}^{6} \xi_j^{(m,n)} \varphi_{jR}^{(m,n)} \tag{4}$$

where $\varphi_{jR}^{(m,n)}$ is the radiated velocity potential induced by the unitary-amplitude $j^{th}$ degree-of-freedom (DOF) motion of the macro-submodule $(m, n)$.

The governing equations of velocity potentials and the corresponding boundary conditions for describing the hydrodynamics of the freely floating multi-rigid-bodies due to the 'macro-submodule division' of the floating flexible structure are given as follows,

$$\begin{cases} \nabla^2 \phi^* = 0 & \text{in } \Omega \\ -\omega^2 \phi^* + g \dfrac{\partial \phi^*}{\partial z} = 0 & \text{on } S_F \\ \dfrac{\partial \phi^*}{\partial z} = 0 & \text{on } S_B \\ \dfrac{\partial (\phi_I + \phi_S)}{\partial z} = 0 & \text{on } \sum S^{(m,n)} \\ \dfrac{\partial \phi_{jR}^{(m,n)}}{\partial n^{(p,q)}} = \begin{cases} v_j^{(m,n)} \cdot \mathrm{n}_j^{(m,n)}, & p=m, q=n \\ 0, & \text{else} \end{cases} & \text{on } S^{(p,q)} \begin{pmatrix} m, p = 1,2,\cdots,M; \\ n, q = 1,2,\cdots,N; j = 1,2,\cdots,6 \end{pmatrix} \\ \lim\limits_{r \to \infty} \sqrt{r} \left( \dfrac{\partial \phi^\times}{\partial r} - \dfrac{i\omega^2}{g} \phi^\times \right) = 0 & \text{on } S_\infty \end{cases} \quad (5)$$

where $\Omega$ is the fluid domain, which is bounded by the free surface $S_F$, the bottom $S_B$, the wetted surface of all macro-submodules $S^{(m,n)}$ $(m = 1,2,\cdots,M; n = 1,2,\cdots,N)$ and the cylindrical surface at infinity $S_\infty$; $v_j^{(m,n)}$ is the $j^{th}$ DOF oscillating velocity of macro-submodule $(m, n)$; $\left( \mathrm{n}_1^{(m,n)}, \mathrm{n}_2^{(m,n)}, \mathrm{n}_3^{(m,n)} \right)^\mathrm{T} = \mathbf{n}^{(m,n)}$ is the unit vector of the macro-submodule $(m, n)$ that is normal to the wetted surface and pointing inward to the body; $\left( \mathrm{n}_4^{(m,n)}, \mathrm{n}_5^{(m,n)}, \mathrm{n}_6^{(m,n)} \right)^\mathrm{T} = \mathbf{s}^{(m,n)} \times \mathbf{n}^{(m,n)}$ with $\mathbf{s}^{(m,n)}$ being the position vector of the body surface of the macro-submodule $(m, n)$; $\phi^*$ can be replaced by $\phi_I$, $\phi_S$ or $\phi_R$; $\phi^\times$ can be replaced by $\phi_S$ or $\phi_R$.

After solving the velocity potentials using the boundary element method, the relevant hydrodynamic coefficients are calculated by

$$\begin{cases} F_{jE}^{(m,n)} = i\omega\rho \iint_{S^{(m,n)}} (\phi_I + \phi_S) \cdot n_j^{(m,n)} dS \\ \omega^2 \xi_k^{(p,q)} A_{j,k}^{[(m,n),(p,q)]} + i\omega \xi_k^{(p,q)} B_{j,k}^{[(m,n),(p,q)]} = \omega^2 \xi_k^{(p,q)} \rho \iint_{S^{(m,n)}} \varphi_{kR}^{(p,q)} \dfrac{\partial \varphi_{jR}^{(m,n)}}{\partial n_j^{(m,n)}} dS \\ \mathbf{C}_j^{[(m,n),(m,n)]} \boldsymbol{\xi}^{(m,n)} = -\rho g \iint_{S^{(m,n)}} \left( \xi_3^{(m,n)} + \xi_4^{(m,n)} y - \xi_5^{(m,n)} x \right) \cdot n_j^{(m,n)} dS \\ m, p = 1,2,\cdots,M; \; n, q = 1,2,\cdots,N; \; j, k = 1,2,\cdots,6 \end{cases} \quad (6)$$

where $F_{jE}^{(m,n)}$ is the $j^{th}$ DOF wave excitation force exerted on the macro-submodule $(m, n)$; $\rho$ the density of water, $\omega$ the wave frequency, $A_{j,k}^{[(m,n),(p,q)]}$ the added mass coefficient in the $j^{th}$ DOF of the macro-submodule $(m,n)$ induced by the $k^{th}$ DOF motion of the macro-submodule $(p,q)$ and $B_{j,k}^{[(m,n),(p,q)]}$ the radiation damping coefficient, $\mathbf{C}_j^{[(m,n),(m,n)]}$ is the $j^{th}$ row of the hydrostatic restoring coefficient matrix $\mathbf{C}^{[(m,n),(m,n)]}$, $g$ the gravitational acceleration, $x$ and $y$ the coordinates of a given point on $S^{(m,n)}$ in the body fixed coordinate system.

Eq. 6 gives the hydrodynamic coefficients including the added mass $\mathbf{A}(\omega)$, the radiation damping $\mathbf{B}(\omega)$, and wave excitation force $\mathbf{F}_E$, as well as the hydrostatic restoring stiffness matrix $\mathbf{C}$, which are elaborated in **Appendix A**. With the defined matrices, the added mass force $\mathbf{F}_A$, the radiation damping force $\mathbf{F}_{Rd}$, the hydrostatic restoring force $\mathbf{F}_{Hs}$ and the inertia force $\mathbf{F}_{In}$ are expressed as

$$\mathbf{F}_A = \omega^2 \mathbf{A}(\omega)\boldsymbol{\xi} \tag{7}$$

$$\mathbf{F}_{Rd} = i\omega \mathbf{B}(\omega)\boldsymbol{\xi} \tag{8}$$

$$\mathbf{F}_{Hs} = -\mathbf{C}\boldsymbol{\xi} \tag{9}$$

$$\mathbf{F}_{In} = \omega^2 \mathbf{M}\boldsymbol{\xi} \tag{10}$$

Together with $\mathbf{F}_E$, the external force acting on all lumped masses $\mathbf{F}_{EXT}$ are defined as

$$\mathbf{F}_{EXT} = \mathbf{F}_E + \mathbf{F}_A + \mathbf{F}_{Rd} + \mathbf{F}_{Hs} + \mathbf{F}_{In} \tag{11}$$

## 2.3 Finite element analysis

Apart from the external force given in Section 2.2, the structural deformation induced force $\mathbf{F}_{St}$ should also be considered for all the lumped masses, which are abstracted from the discretized macro-submodules. The structural force can be calculated by the lumped-mass stiffness matrix with respect to all lumped masses, $\mathbf{K}$ and the unknown displacements of lumped masses, $\boldsymbol{\xi}$ as follows,

$$\mathbf{F}_{St} = -\mathbf{K}\boldsymbol{\xi} \tag{12}$$

As a result, the key to calculating the structural force is the determination of the lumped-mass stiffness matrix, the procedure of which is shown in Fig. 5. It is emphasized that the discretization strategies presented in Fig. 5 only serve to give a better illustration (similarly hereinafter). A convergence study is needed to finalize the discretization strategy, which will be dealt with in Section 4.1.2.

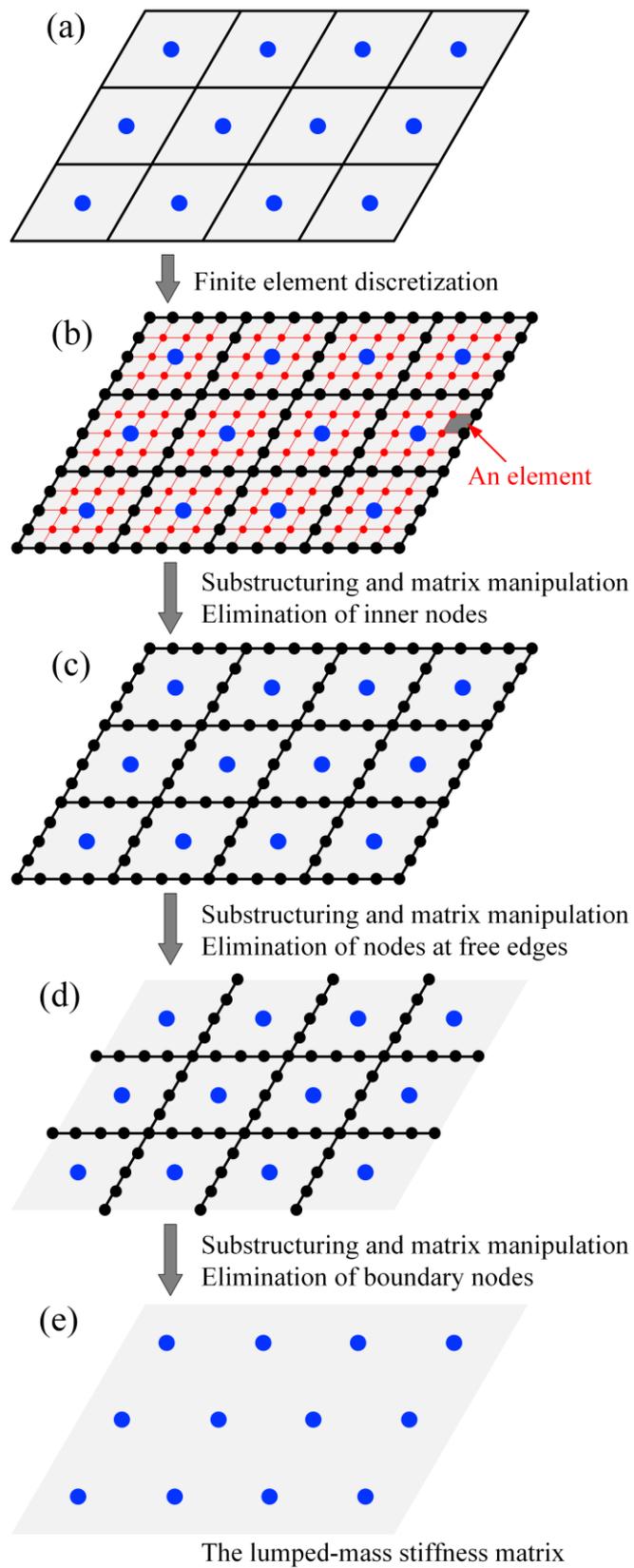

Fig. 5. The strategy to derive the lumped-mass stiffness matrix. The blue solid circles are the lumped masses, the black ones are the nodes at boundaries of each macro-submodule and the red ones are the nodes inside each macro-

submodule. It is noted that the discretization strategies shown here only serve to give a better illustration and understanding. Convergence study is needed to finalize the discretization strategy.

### 2.3.1 Finite element discretization

As indicated by Eq. 11, all distributed loads acting on a certain macro-submodule have now been transformed to concentrated external forces acting on the corresponding lumped mass, which means the floating flexible structure is subjected to external forces only on points where lumped masses located (see Fig. 5a). Consequently, as shown in Fig. 6, the macro-submodule $(m,n)$ is subjected to external forces $\mathbf{F}_{EXT}^{(m,n)}$ on the lumped mass (blue solid circle) and boundary force on boundaries (black solid lines). Here, the boundary force is actually the internal force which is induced by the structure's elastic deformation and exposed when the macro-submodule is cut (or divided) from the floating flexible structure. No force is exerted elsewhere.

Finite element discretization is adopted. Since all macro-submodules are identical, the same discretization strategy is used, which is shown in Fig. 6. The discretization is performed in the body-fixed coordinate system. A four-node element with 6 DOF at each node is adopted, that is, translations in the $x_{(m,n)}$, $y_{(m,n)}$ and $z_{(m,n)}$ directions and rotations about the $x_{(m,n)}$, $y_{(m,n)}$ and $z_{(m,n)}$ axes. It is noted that a lumped mass coincides with one of the nodes for the convenience of derivation of the lumped-mass stiffness matrix. As shown in Fig. 6, all nodes are categorized into three kinds: the lumped mass (blue solid circles), boundary nodes (black ones) and inner nodes (red ones). Clearly, the lumped mass is where external force is exerted, boundary nodes are where boundary force is exerted and no force is exerted on inner nodes. Introduction of finite element discretizes the structure from Fig. 5a to Fig. 5b.

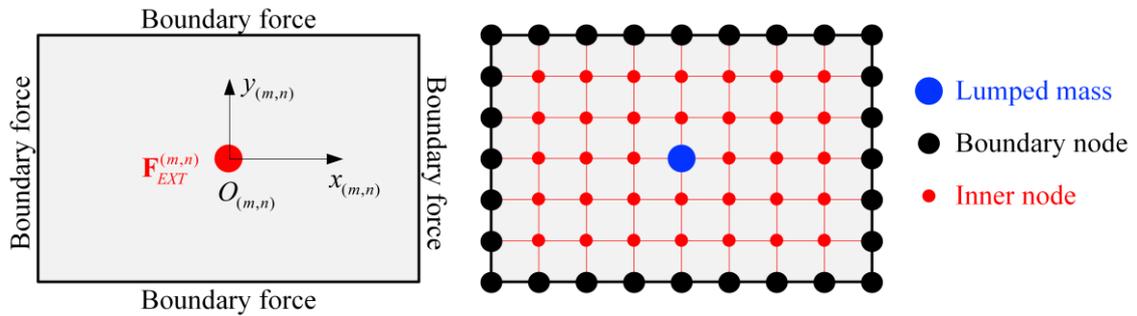

**Fig. 6. Finite element discretization on the macro-submodule. This macro-submodule is cut from the original structure with external force acting on the lumped mass (blue solid circles) and boundary force acting on the boundaries (black solid lines). The boundary force is the internal force induced by elastic deformation and exposed when a macro-submodule is cut from the original structure.**

### 2.3.2 Substructuring and matrix manipulation

The stiffness matrix of all nodes related to a given macro-submodule in Fig. 6 could be derived based on the finite element theory. As the finite element theory is rather mature and common, derivation of this stiffness matrix is not presented in this paper, and main focuses are put on the DMFE method. Extra attention is needed in node numbering, which underlies the form of this stiffness matrix. Recommended numbering rules are given: the lumped mass is always numbered as '1', followed by all boundary nodes and then the inner nodes. Denote $\mathbf{k}_{\text{module}}$ as the stiffness matrix of all nodes in Fig. 6 to relate displacement and force:

$$\mathbf{k}_{\text{module}} \begin{bmatrix} \boldsymbol{\xi}^{(m,n)}_{\text{lumped-mass}} \\ \boldsymbol{\xi}^{(m,n)}_{\text{boundary}} \\ \boldsymbol{\xi}^{(m,n)}_{\text{inner}} \end{bmatrix} = \begin{bmatrix} \mathbf{F}^{(m,n)}_{EXT} \\ \mathbf{F}^{(m,n)}_{\text{boundary}} \\ \mathbf{0} \end{bmatrix} \tag{13}$$

where $\boldsymbol{\xi}^{(m,n)}_{\text{lumped-mass}}$ is the displacement of lumped mass $(m,n)$, $\boldsymbol{\xi}^{(m,n)}_{\text{boundary}}$ is the displacement of all boundary nodes on macro-submodule $(m,n)$, $\boldsymbol{\xi}^{(m,n)}_{\text{inner}}$ is the displacement of all inner nodes on macro-submodule $(m,n)$ and $\mathbf{F}^{(m,n)}_{\text{boundary}}$ is the force acting on all boundary nodes of macro-submodule $(m,n)$, $\mathbf{0}$ indicates no force is exerted on the inner nodes.

It is worth noting that $\mathbf{k}_{\text{module}}$ is already a matrix of huge size, let alone the stiffness matrix of the whole structure (containing all nodes together) after combining $\mathbf{k}_{\text{module}}$ of every macro-submodule. Based on the fact that zero force is exerted on all the inner nodes, some manipulations could be performed to derive the following equations after 'eliminating' the inner nodes:

$$\boldsymbol{\xi}^{(m,n)}_{\text{inner}} = -\boldsymbol{\Lambda}_{\text{INNER}} \begin{bmatrix} \boldsymbol{\xi}^{(m,n)}_{\text{lumped-mass}} \\ \boldsymbol{\xi}^{(m,n)}_{\text{boundary}} \end{bmatrix} \tag{14}$$

$$\begin{bmatrix} \mathbf{F}^{(m,n)}_{EXT} \\ \mathbf{F}^{(m,n)}_{\text{boundary}} \end{bmatrix} = \mathbf{K}_{\text{OUTER}} \begin{bmatrix} \boldsymbol{\xi}^{(m,n)}_{\text{lumped-mass}} \\ \boldsymbol{\xi}^{(m,n)}_{\text{boundary}} \end{bmatrix} \tag{15}$$

where a detailed derivation and the form of $\boldsymbol{\Lambda}_{\text{INNER}}$ and $\mathbf{K}_{\text{OUTER}}$ are given in **Appendix B**.

$\mathbf{K}_{\text{OUTER}}$ in Eq. 15 directly describes the relationship between the force and the displacement related with the lumped-mass and boundary nodes of a given macro-submodule, which takes the structure from Fig. 5b to Fig. 5c, where all the inner nodes are eliminated.

It is noted that extra manipulation is needed on $\mathbf{K}_{\text{OUTER}}$ for some macro-submodules that have boundary nodes at free edges of the floating flexible structure (see Fig. 7a, highlighted in dark grey). As an example, boundary nodes of macro-submodule $(1,1)$ framed in red dashed box are the nodes at free edges (Fig. 7b), which are subjected to no boundary force. An exception is the nodes $j$ and $k$ at the free edges as they both belong to the interface of two different macro-submodules and the cutting (or division) operation exposes the nodes to non-zero forces. Based on the above analysis on macro-submodule $(1,1)$, for all macro-submodules with boundary nodes at free edges (see Fig. 7a), the following equation is drawn

$$\mathbf{K}_{\text{OUTER}} \begin{bmatrix} \boldsymbol{\xi}^{(m,n)}_{\text{lumped-mass}} \\ \boldsymbol{\xi}^{(m,n)}_{\text{non-free}} \\ \boldsymbol{\xi}^{(m,n)}_{\text{free-edge}} \end{bmatrix} = \begin{bmatrix} \mathbf{F}^{(m,n)}_{EXT} \\ \mathbf{F}^{(m,n)}_{\text{non-free}} \\ \mathbf{0} \end{bmatrix} \tag{16}$$

Here, the boundary nodes are categorized into non-free-edge nodes (non-free in short) and free-edge nodes with boundary force $\mathbf{F}^{(m,n)}_{\text{non-free}}$ exerted on the former and no force exerted on the latter, which explains the vector '$\mathbf{0}$' in Eq. 16.

Similarly, the substructuring and some manipulations are performed to 'eliminate' all the free-edge nodes:

$$\xi_{\text{free-edge}}^{(m,n)} = -\Lambda_{\text{FREE-EDGE}} \begin{bmatrix} \xi_{\text{lumped-mass}}^{(m,n)} \\ \xi_{\text{non-free}}^{(m,n)} \end{bmatrix} \quad (17)$$

$$\begin{bmatrix} F_{EXT}^{(m,n)} \\ F_{\text{non-free}}^{(m,n)} \end{bmatrix} = K_{\text{NON-FREE}} \begin{bmatrix} \xi_{\text{lumped-mass}}^{(m,n)} \\ \xi_{\text{non-free}}^{(m,n)} \end{bmatrix} \quad (18)$$

where a detailed derivation and the form of $\Lambda_{\text{FREE-EDGE}}$ and $K_{\text{NON-FREE}}$ are given in **Appendix C.**

$K_{\text{NON-FREE}}$ in Eq. 18 directly describes the relationship between the force and the displacement related with the lumped-mass and non-free-edge nodes of a given submodule, which takes the structure from Fig. 5c to Fig. 5d, where all the boundary nodes exposed to zero boundary force are eliminated.

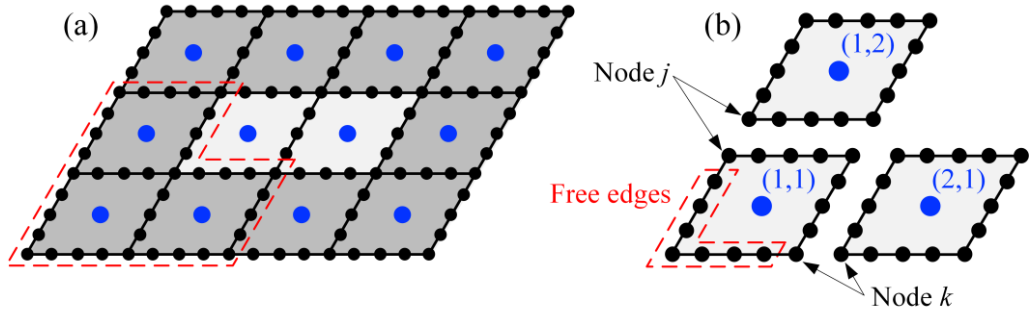

**Fig. 7. (a) Illustrations of macro-submodules with nodes at free edges. (b) illustration of free-edge nodes on macro-submodule $(\mathbf{1}, \mathbf{1})$ where no boundary forces are exerted. Despite at the free edge, boundary force exists on Node $j$ and Node $k$.**

It is emphasized that $\Lambda_{\text{FREE-EDGE}}$ and $K_{\text{NON-FREE}}$ differ among the 'dark grey' macro-submodules shown in Fig. 7a for the free-edge nodes appear differently in each macro-submodule, thus calculation is needed on each of them. On the contrary, $\Lambda_{\text{INNER}}$ and $K_{\text{OUTER}}$ are the same for each macro-submodule.

Another numbering is needed on all nodes shown in Fig. 5d, including all the lumped-masses and all the boundary nodes remained. Recommended rules are the lumped masses being numbered first and then the remained boundary nodes. Fig. 8 gives all nodes of each macro-submodule that are not eliminated with blue solid circles being the lumped masses, black ones being the remained boundary nodes and the translucent ones being the eliminated boundary nodes (shown here only for better illustration). The stiffness matrix of macro-submodules without free-edge nodes is calculated by Eq. 15 and that of macro-submodules with free-edge nodes are calculated by Eq. 18. With the node numbering, the structural stiffness matrix of Fig. 5d is constructed based on $K_{\text{OUTER}}$ and $K_{\text{NON-FREE}}$ following the standard finite element procedure, which is denoted $K_{\text{ALL}}$.

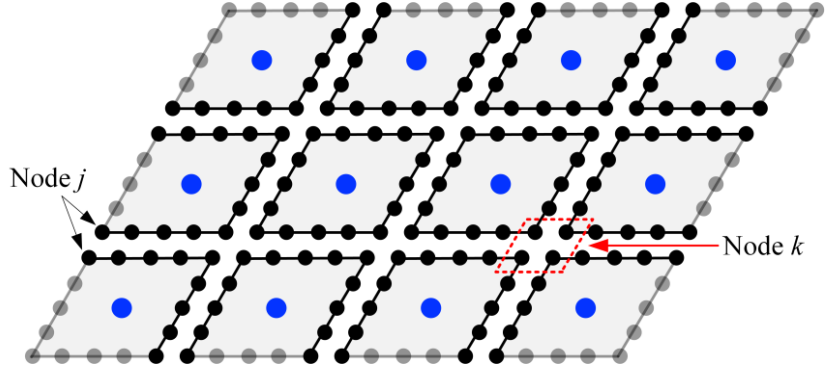

**Fig. 8. All nodes remained in each macro-submodule where force acts. The blue solid circles are the lumped masses, the black ones are the boundary nodes where the boundary force is exerted and the translucent ones are the eliminated free-edge nodes where no boundary force exists.**

Every node given in Fig. 8, the blue ones and the translucent ones aside, must appear in at least two macro-submodules, which ensures the existence of boundary force. When the stiffness matrix $\mathbf{K}_{ALL}$ is constructed, boundary force acting on each node will be added accordingly. More importantly, the sum of all boundary forces corresponding to a given node is undoubtedly zero. For instance, Node $k$ in Fig. 8 appears in four macro-submodules and in each macro-submodule a boundary force is exerted on Node $k$. The sum of the four boundary forces is zero. Similarity is observed for Node $j$ with the only difference being this node appears in two macro-submodules. Above analysis gives

$$\begin{bmatrix} \mathbf{F}_{St} \\ \mathbf{0} \end{bmatrix} = \begin{bmatrix} -\mathbf{F}_{EXT} \\ \mathbf{0} \end{bmatrix} = -\mathbf{K}_{ALL} \begin{bmatrix} \boldsymbol{\xi} \\ \boldsymbol{\xi}_{rest} \end{bmatrix} = -\begin{bmatrix} \mathbf{K}_{ALL}^{1,1} & \mathbf{K}_{ALL}^{1,2} \\ \mathbf{K}_{ALL}^{2,1} & \mathbf{K}_{ALL}^{2,2} \end{bmatrix} \begin{bmatrix} \boldsymbol{\xi} \\ \boldsymbol{\xi}_{rest} \end{bmatrix} \tag{19}$$

where $\mathbf{K}_{ALL}^{j,k}(j,k=1,2)$ are the sub-matrices of $\mathbf{K}_{ALL}$ and $\boldsymbol{\xi}_{rest}$ is the displacement of all black nodes in Fig. 5d or Fig. 8. $\mathbf{F}_{St}$ is the structural deformation induced force on the lumped masses, which together with the external force $\mathbf{F}_{EXT}$, leads to the equilibrium condition following the d'Alembert principle. The meaning of $\mathbf{F}_{EXT}$ is given in Eq. 11 and that of $\boldsymbol{\xi}$ in Eq. A6.

Some manipulations give

$$\boldsymbol{\xi}_{rest} = -[\mathbf{K}_{ALL}^{2,2}]^{-1}\mathbf{K}_{ALL}^{2,1}\boldsymbol{\xi} \triangleq -\boldsymbol{\Lambda}_{REST} \cdot \boldsymbol{\xi} \tag{20}$$

$$\mathbf{F}_{St} = -\mathbf{F}_{EXT} = -\{\mathbf{K}_{ALL}^{1,1} - \mathbf{K}_{ALL}^{1,2}\boldsymbol{\Lambda}_{REST}\}\boldsymbol{\xi} \triangleq -\mathbf{K} \cdot \boldsymbol{\xi} \tag{21}$$

$\mathbf{K}$ in Eq. 21 is the targeted lumped-mass stiffness matrix that directly describes the relationship between the structural force acting on the lumped masses and the displacement of them, which takes the structure from Fig. 5d to Fig. 5e, where only the lumped masses are reserved.

### 2.3.3 The hydroelastic equation

Eqs. 7-11 and 21 give the hydroelastic equation that the DMFE method establishes on all the lumped masses

$$\{-\omega^2(\mathbf{M}+\mathbf{A}(\omega))-i\omega\mathbf{B}(\omega)+(\mathbf{C}+\mathbf{K})\}\boldsymbol{\xi}=\mathbf{F}_E \tag{22}$$

Solution of Eq. 22 is the displacement response of the lumped masses $\boldsymbol{\xi}$.

## 2.4 Recovery of displacement response

After obtaining the displacements at the positions of lumped masses, displacement response (at any given position) of the whole structure is solved reversely compared to the derivation of the lumped-mass matrix. First, the deformation of all nodes in Fig. 5d is obtained through Eq. 20 with $\boldsymbol{\xi}$ solved from Eq. 22, the displacement response of all the lumped masses, i.e., all nodes in Fig. 5e. Then, that of all the translucent nodes in Fig. 8 is solved through Eq. 17, after which the displacement response of all boundary nodes and the lumped mass is known for each macro-submodule, which takes the procedure to Fig. 5c. Finally, Eq. 14 gives the displacement of all the inner nodes and finalizes the recovery of displacement response.

It can be seen that, although finite element is introduced, the DMFE method simplifies the stiffness matrix of the whole structure, a surprisingly huge matrix, to the small-sized lumped-mass stiffness matrix by substructuring and matrix manipulations, which improves the computation efficiency significantly. Besides, as one of the unique features, modal analysis is not needed in the DMFE method.

## 2.5 Recovery of internal force response

The essence of calculating internal force response is integration of external loads, i.e., hydrodynamic pressure. Both the mode-superposition method and the direct method is able to obtain the 'real-case' hydrodynamic pressure distribution as the structure is considered elastic. However, all macro-submodules are considered rigid when the DMFE method performs hydrodynamic analysis, which means the DMFE method cannot give the actual hydrodynamic pressure distribution.

Definitions are regulated first before describing the recovery strategies. For a flexible floating structure, whose horizontal dimensions is much larger than the vertical one, bending moment is a primary concern in safety assessment. Two different bending moments appear in this paper, i.e., $M_y$ (revolving around the $Y$ axis) and $M_x$ (revolving around the $X$ axis). It is emphasized that all internal forces presented in this paper are values per unit width.

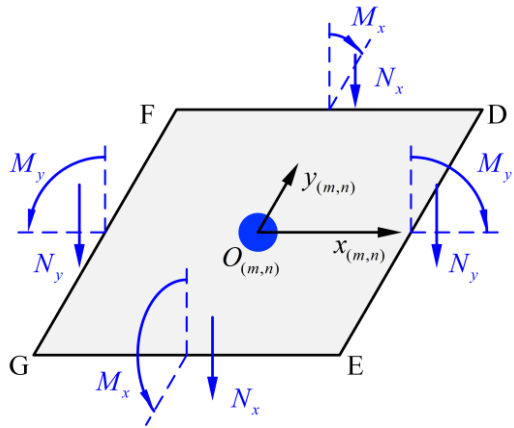

**Fig. 9.** Bending moment and shear force acting on macro-submodule $(m, n)$ with four vertexes $E, D, F, G$. Two different bending moments are defined, i.e., $M_y$ (revolving around the $Y$ axis) and $M_x$ (revolving around the $X$ axis). The shear force on edges normal to the $x_{(m,n)}$ axis is denoted as $N_y$ and that on edges normal to the $y_{(m,n)}$ axis is denoted as $N_x$.

Though hydrodynamic pressure is unsolvable in the framework of the DMFE method, the internal (boundary) force is 'exposed' when a macro-submodule is cut (divided) from the original structure (see Fig. 6). Take macro-submodule $(m, n)$ colored dark grey in Fig. 3 as an example, whose finite element distribution is shown in Fig. 6 and internal force illustrated in Fig. 9 with $E, D, F, G$ being the four vertexes. With displacement response of the whole structure solved in Section 2.4, Eq. 15 gives the internal force acting on all boundary nodes located in edges $ED, DF, FG, GE$, each being a six DOF vector, which is denoted as $F_i (i = 1, 2, \cdots, 6)$. Fig. 10a gives the six-component internal force vector of boundary nodes on edge $DF$. It can be seen combined with Fig. 9, that the third component $F_3$ corresponds to $N_x$ in Fig. 9 and the fourth component $F_4$ corresponds to $M_x$. Boundary nodes on edge $GE$ share the same features with those on edge $DF$. Fig. 10b gives the six-component internal force vector of boundary nodes on edge $DE$. It can be seen combined with Fig. 9, that the third component $F_3$ corresponds to $N_y$ in Fig. 9 and the fifth component $F_5$ corresponds to $M_y$. Boundary nodes on edge $GF$ share the same features with those on edge $DE$ In conclusion, the DMFE method is able to obtain the internal force of all boundary nodes (interfaces), of which only the following enters the bending moment recovery: shear force $N_y$ and bending moment $M_y$ of boundary nodes located on edges normal to the $O_{(m,n)} - x_{(m,n)}$ axis; shear force $N_x$ and bending moment $M_x$ of boundary nodes located on edges normal to the $O_{(m,n)} - y_{(m,n)}$ axis.

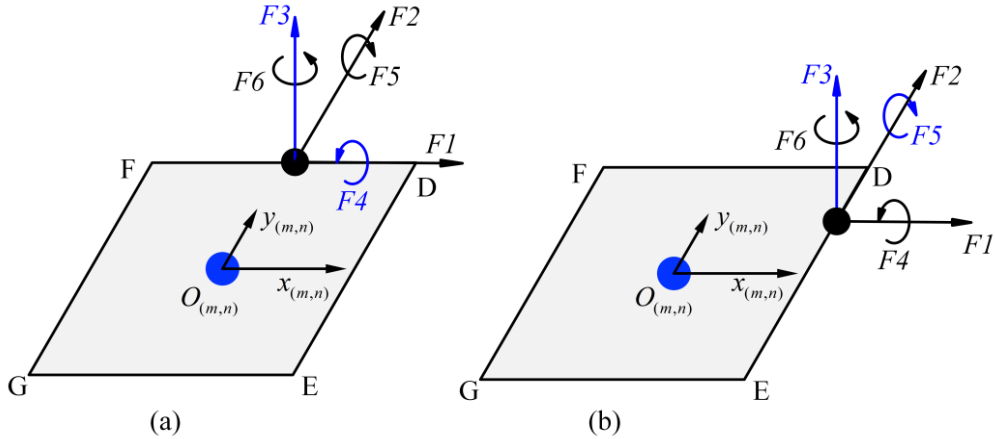

**Fig. 10.** Illustration of the six-component force vector of nodes on edge $DF$ (a) and edge $DE$ (b). In (a), the third component $F_3$ corresponds to $N_x$ in Fig. 9 and the fourth component $F_4$ corresponds to $M_x$. Similarly, in (b), $F_3$ corresponds to $N_y$ in Fig. 9 and $F_5$ corresponds to $M_y$.

In fact, the finite element theory will also give the 'internal force' on every inner node (red solid circles in Fig. 6). As shown in Fig. 11, one can cut the macro-submodule $(m, n)$ in Fig. 6 and expose the 'internal force' on corresponding inner nodes. This 'internal force' is denoted as exposed force $\mathbf{F}_{\text{exposed}}$. The stiffness matrix of the dark grey structure $\mathbf{K}_{\text{exposed}}$ can be constructed and the deformation of every node $\boldsymbol{\xi}_{\text{exposed}}$ is known. The product of the two terms will give forces acting on each node, $\mathbf{F}_{\text{exposed}}$ included apparently. Notice that the force distribution on a given macro-submodule has changed since all external loads are concentrated on the lumped mass, in which case, the derived $\mathbf{F}_{\text{exposed}}$ does not reflect the internal force response of the original loading case. However, the 'force concentration' procedure imposes no effect on the boundary force (the internal force of nodes on interfaces) of a given macro-submodule. This indicates that, in the framework of the DMFE method, the recovery of internal force could only be done through some interpolation scheme using the 'correct' boundary force of the macro-submodule as a basis, which is one of the distinctive features of the DMFE method.

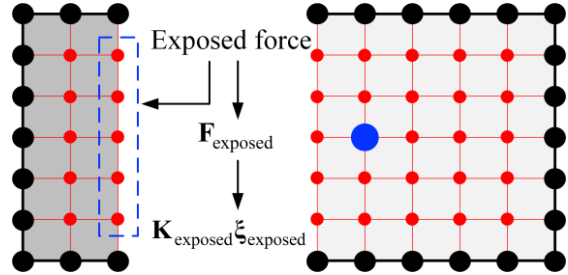

**Fig. 11. Macro-submodule $(m, n)$ in Fig. 6 is cut (divided) into two parts, in which case the internal force acting on inner nodes framed in blue dashed box will be exposed, which is denoted as exposed force $\mathbf{F}_{\text{exposed}}$. The stiffness matrix of the exposed part (colored in dark grey) can be constructed (denoted as $\mathbf{K}_{\text{exposed}}$) and the deformation of all nodes in the dark grey part can be solved in Section 2.4 (denoted as $\boldsymbol{\xi}_{\text{exposed}}$). Apparently, $\mathbf{F}_{\text{exposed}} = \mathbf{K}_{\text{exposed}} \boldsymbol{\xi}_{\text{exposed}}$.**

Strategies to calculate the bending moment distribution $M_y(x, y)$ is presented. Above analysis gives that only nodes located on the edges normal to $O_{(m,n)} - x_{(m,n)}$ axis (see the solid circles in Fig. 12) enter the calculation. Particularly, the ones framed in the blue dashed box are located at the free ends with zero internal force ($M_y = N_y = 0$).

The red solid line in Fig. 12 ($y = y_0$) penetrates through a certain number of interfaces, which means the bending moment values are known on certain specific locations (boundary nodes, i.e., solid circles). The spline with zero slope at free ends is used to connect all values. This drawn spline curve is then the bending moment distribution on this specific position $M_y(x, y)|_{y=y_0}$. Repetition of above step on every position will give the bending moment distribution $M_y(x, y)$ along the structure. Identical strategies and procedures are used to calculate $M_x(x, y)$, with the only difference being the red solid line in Fig. 12 is drawn parallel to the $Y$ axis. This strategy is named as 'the spline method'.

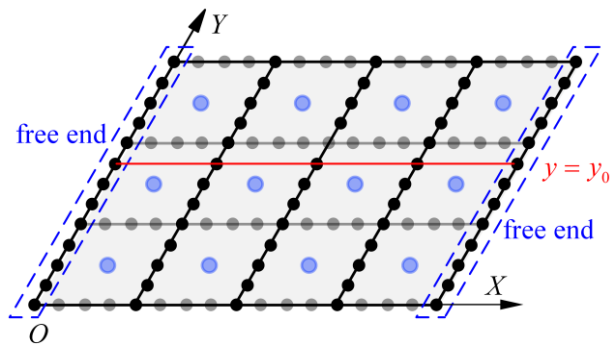

**Fig. 12. Illustration of the spline method. Solid circles are the boundary nodes that enters the calculation of bending moment $M_y(x, y)$. The ones framed in the dashed blue box are located at the free ends with zero internal**

force ($M_y = N_y = 0$). $M_y$ and $N_y$ of the black ones are calculated through Eq. 15. A spline with zero slope at free ends is used to connect all bending moment values on boundary nodes (solid circles on the red line). This drawn spline is then the bending moment distribution $M_y(x, y)\big|_{y=y_0}$

## 3 Relations with the DMB method

The DMFE method and the DMB method differ mainly in two aspects, that is, macro-submodule division and derivation of the lumped-mass stiffness matrix (see Fig. 13.), which limit the DMB method's implementation only on VLFSs with a large length/width ratio.

The DMB method divides the structure into macro-submodules only longitudinally, which is reasonable as the targeted structure is long and narrow. Lumped masses lie identically at each macro-submodule's center of gravity. Finite element discretization is not adopted to derive the lumped-mass stiffness matrix. Instead, beam elements are introduced to connect every adjacent lumped mass, which leaves a beam-connected-lumped-masses system. The beam element is abstracted from the macro-submodule between adjacent lumped masses (framed in red dashed box) and inherits the physical and mechanical properties of the original structure. Since the theoretical solution of a beam element's structural stiffness matrix exists and the lumped masses are numbered, the lumped-mass stiffness matrix could be directly obtained by installing the stiffness matrix of every beam element following the standard procedure of the finite element method.

Therefore, hydroelasticity is solved in the DMB method but only on the centerline of the structure where the lumped masses and beam elements are located. However, the DMFE method not only undertakes the submodule division in both length and width direction, but discretizes each macro-submodule with finite elements that cover the whole structure. The lumped-mass stiffness matrix is derived through substructuring and matrix manipulations, which assures that hydroelasticity could be solved everywhere along the large flexible floating structure. Details on the DMB method could be referred to [9-11].

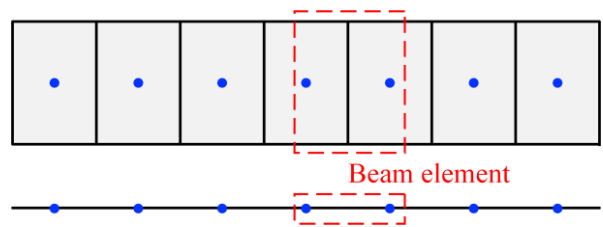

**Fig. 13. Macro-submodule division and introduction of beam element in the DMB method.**

Notice that the DMB method is viewed as a 'one-dimensional' method only results from the one-dimensional macro-submodule division. It is emphasized the following aspects are all three-dimensional, including the VLFS itself, the potential-flow theory and the beam elements adopted.

## 4 Validation and application

### 4.1 A narrow VLFS

The hydroelastic response of the VLFS 'MF-300' under regular waves is studied both experimentally and numerically by numerous researchers [19-22]. Its characteristics are shown in Table. 1. The VLFS remains static before subjected to regular waves coming in from four directions shown in Fig. 14. The red dashed lines indicate three specific positions of the structure and are labeled as 'P' (portside), 'C' (centerline) and 'S' (starboard), respectively. The incident wave amplitude takes the value of 1m in all simulations presented in this paper.

It is emphasized that all hydroelastic response presented in this paper is given in the form of amplitudes as the hydroelastic analysis is performed in the frequency domain. Therefore, amplitudes of vertical displacement (bending moment) are written as vertical displacement (bending moment) for brevity.

**Table 1. Principal particulars of 'MF-300' (Prototype)**

| | |
|---|---|
| Length (m) | 300.0 |
| Breadth (m) | 60.0 |
| Draft (m) | 0.5 |
| Depth (m) | 2.0 |
| Vertical bending stiffness (N·m$^2$) | 4.77e11 |
| Water depth (m) | 58.5 |
| Mass (kg) | 9.225e6 |

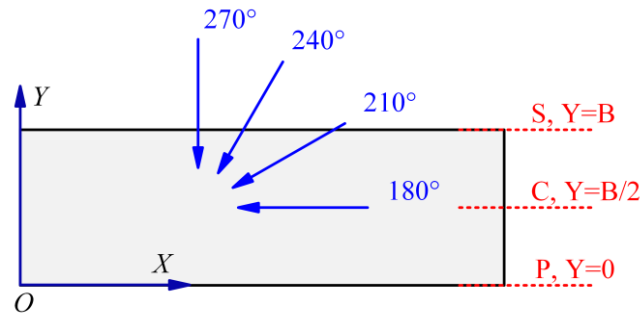

**Fig. 14. Illustration of head (180°), beam (270°) and oblique waves (210° and 240°). Note that this definition contradicts with that defined in [21, 23] (the former is 180° larger than the latter). For example, 0° shown here corresponds to 180° in [21, 23]. Three specific locations are defined, that is, P (Y=0), C (Y=B/2) and S (Y=B).**

### 4.1.2 Convergence study

4.1.2.1 On the macro-submodule division

Convergence study on the macro-submodule division is given first. It has been proven that the hydroelasticity of MF-300 converges when the structure is divided into more than 5 macro-submodules longitudinally [9]. In this paper, MF-300 is divided into 8 macro-submodules in the $X$ direction. Hence, emphasis is placed on the macro-submodule division in the $Y$ direction. Table 2 gives the details of the five different submodule-division strategy with the structure being divided in the $Y$ direction into 1, 2, 3, 4 and 5 parts. Notice the length and the height of each macro-submodule, regardless of the division strategy, are 37.5m and 2m, respectively. Therefore, only the width is given in Table 2. Details on the finite element discretization on each macro-submodule in different strategies are presented in Fig. 15. The square grid is adopted in the finite element discretization and a grid size of 5m is used for all strategies listed in Table 2. It is worth noting that adjustment is made on the grid size near a lumped mass to ensure that the lumped mass coincides with a certain node.

Table 2. Details of strategies that enters the convergence study on the macro-submodule division in the $Y$ direction

| Number of macro-submodules in the $Y$ direction | Number of macro-submodules in the $X$ direction | Size of each macro-submodule (only width is shown) | Finite element discretization |
|---|---|---|---|
| 1 | 8 | 60m | Fig. 15a |
| 2 | 8 | 30m | Fig. 15b |
| 3 | 8 | 20m | Fig. 15c |
| 4 | 8 | 15m | Fig. 15d |
| 5 | 8 | 12m | Fig. 15e |

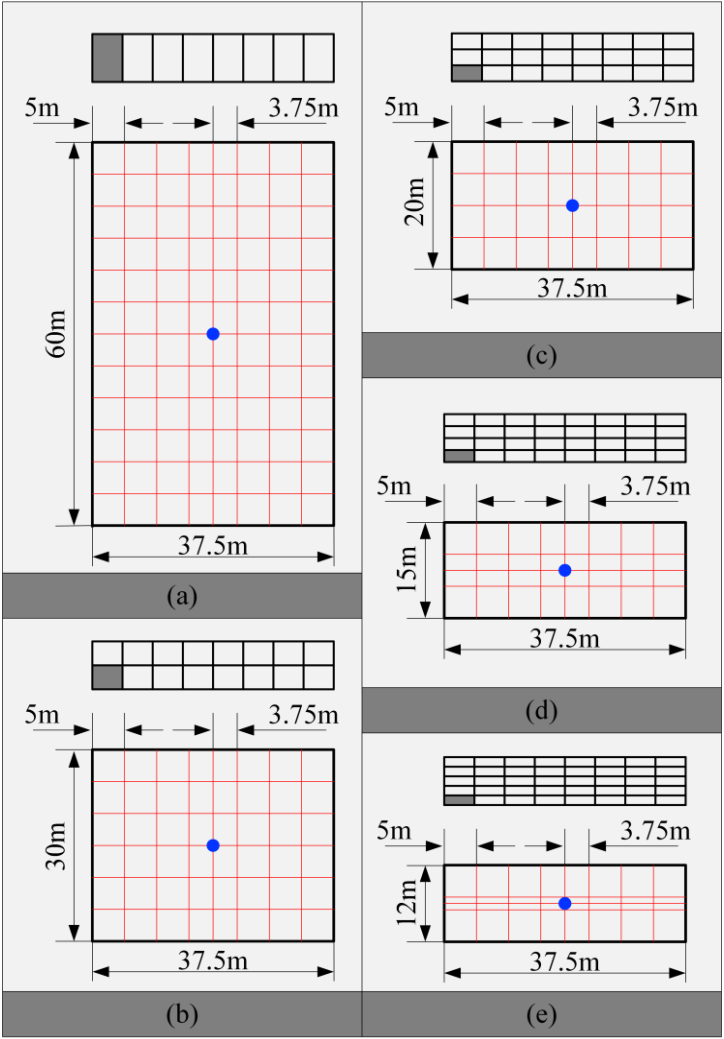

Fig. 15. Finite element discretization on each macro-submodule (Top view) in different division strategies listed in Table 2.

Presented in Fig. 16 are the vertical displacement of MF-300 under a 180m regular wave for four incident angles indicated in Fig. 14.

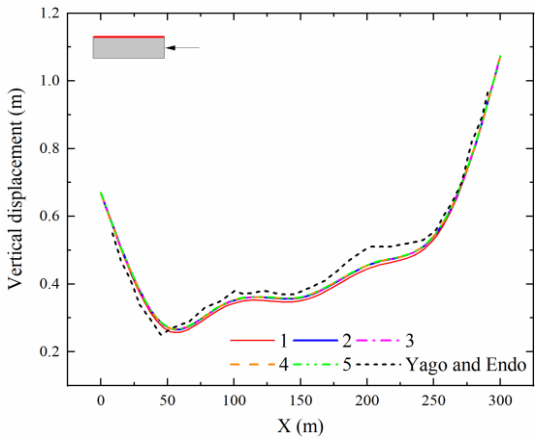
(a) 180m 180degree-S

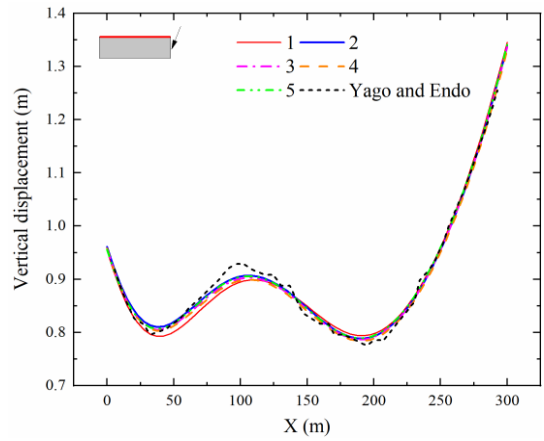
(g) 180m 240degree-S

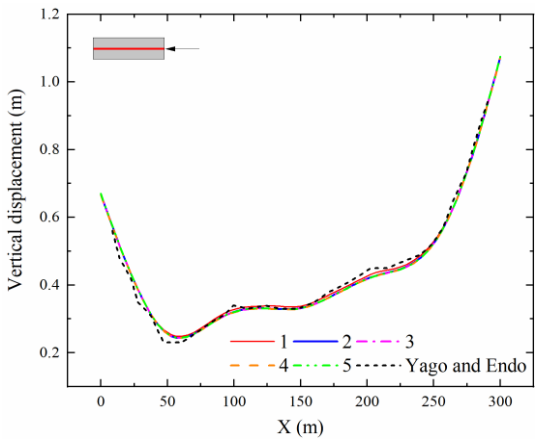
(b) 180m 180degree-C

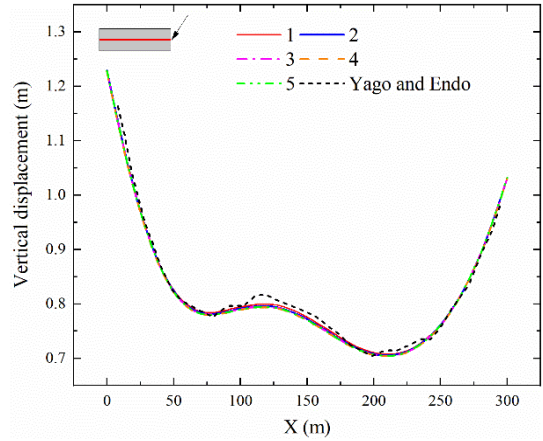
(h) 180m 240degree-C

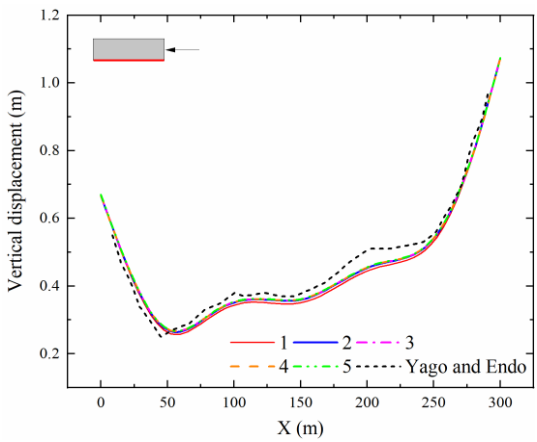
(c) 180m 180degree-P

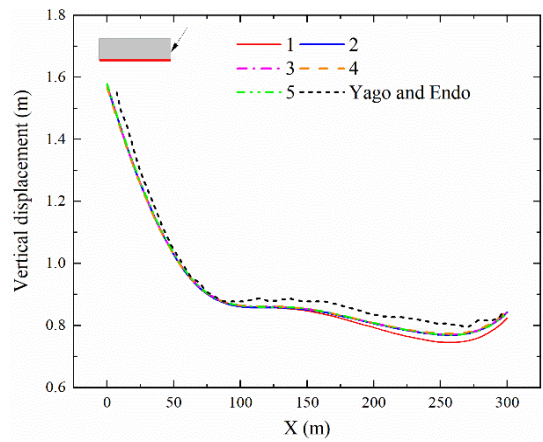
(i) 180m 240degree-P

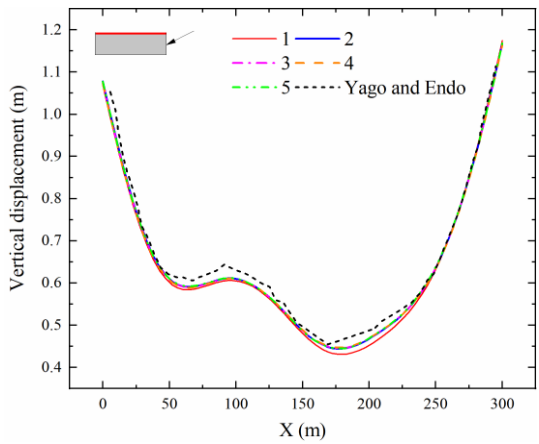

(d) 180m 210degree-S

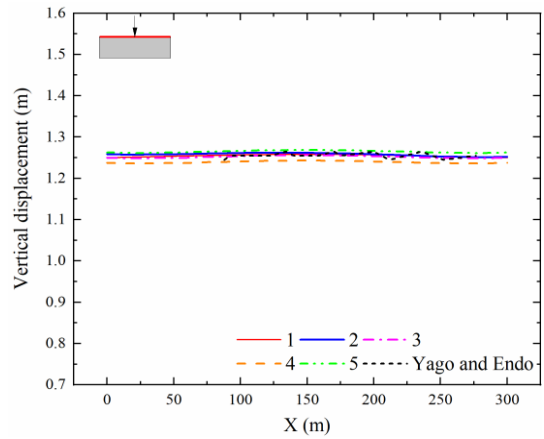

(j) 180m 270degree-S

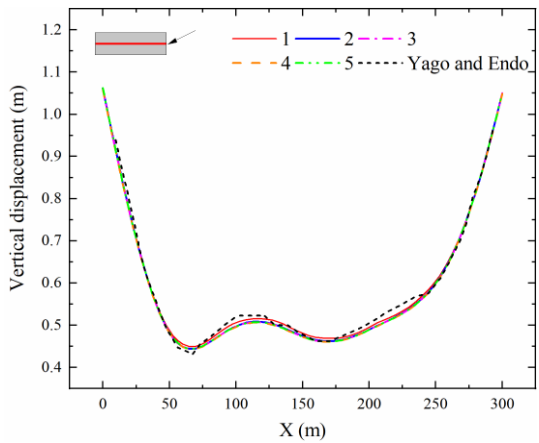

(e) 180m 210degree-C

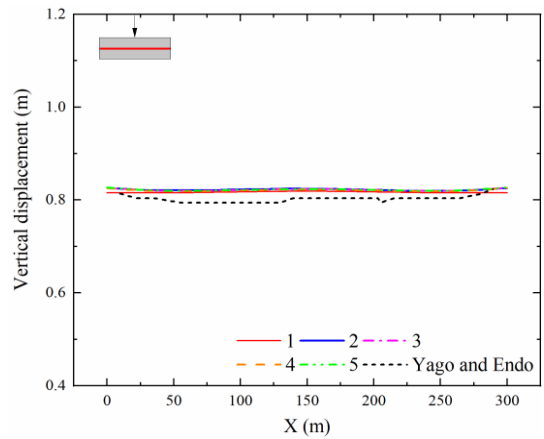

(k) 180m 270degree-C

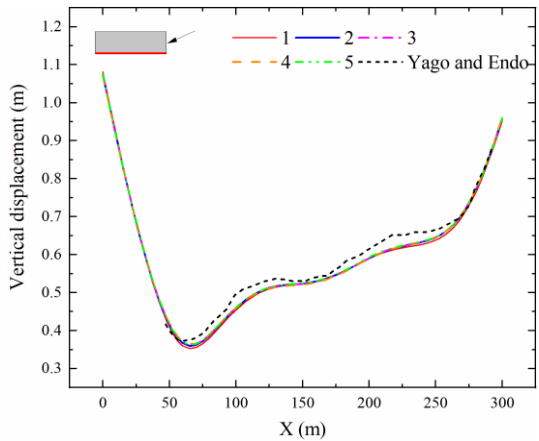

(f) 180m 210degree-P

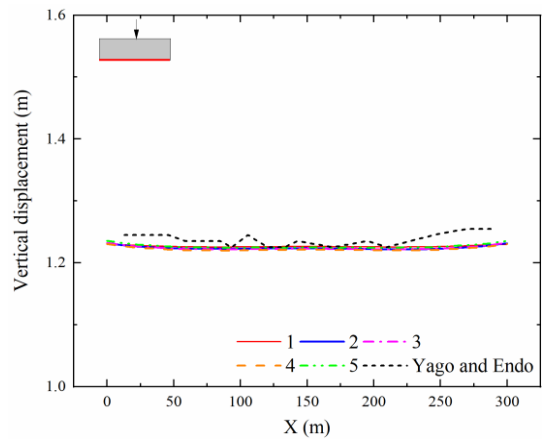

(l) 180m 270degree-P

**Fig. 16. Vertical displacement of MF-300 under a 180m regular wave for four incident angles (180°, 210°, 240° and 270°). Simulations of Yago and Endo [19] using the mode-superposition method is shown in black dashed lines as comparison. Numbers in the legend reflect the macro-submodule division in the $Y$ direction (see Table 2). A grey rectangular that represents MF-300 is located at the up-left corner of each sub-figure with the red solid line indicating the position and the arrow indicating the wave incident angle.**

Firstly, all simulations achieve a rather good agreement with Yago and Endo's simulations, which confirms the accuracy of the DMFE method. Secondly, different macro-submodule division strategies in Table 2 give almost identical results, which means the strategy with one macro-submodule in the $Y$ direction, i.e., no division is done transversely, is capable enough to capture the elastic deformation in the width direction. It is reasonable to draw that the result converges when MF-300 is divided into more than 1 macro-submodule in the $Y$ direction.

4.1.2.2 On the finite element discretization

This section serves to explore the effects of the grid size. The convergence study in this section is performed on the structure with eight macro-submodules divided in the $X$ direction and three in the $Y$ direction. Similarly, the square grid is used. Table 3 gives the five different grid sizes taken, i.e., 1m, 2m, 3m, 4m, and 5m. Fig. 17a illustrates the macro-submodule division strategy and Figs. 17b-f the detailed discretization that correspond to Table 3.

Table 3. Details of strategies that enters the convergence study on the finite element discretization with different grid sizes

| Grid size | Number of macro-submodule in the $Y$ direction | Number of macro-submodule in the $X$ direction | Finite element discretization |
| --- | --- | --- | --- |
| 1m | 3 | 8 | Fig. 17b |
| 2m | 3 | 8 | Fig. 17c |
| 3m | 3 | 8 | Fig. 17d |
| 4m | 3 | 8 | Fig. 17e |
| 5m | 3 | 8 | Fig. 17f |

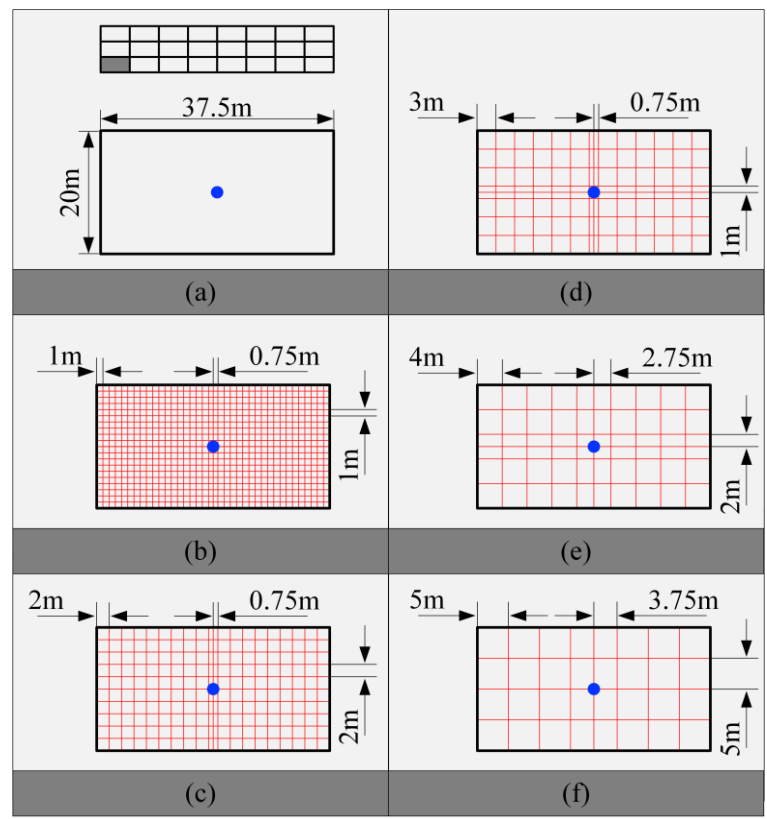

**Fig. 17. Finite element discretization (Top view) on each macro-submodule with different grid sizes listed in Table 3.**

Vertical displacement of MF-300 under a 180m regular wave for four incident angles are given in Fig. 18. Similarly, little difference is observed on the displacement response of strategies with different grid size and all simulations agree well with Yago and Endo's results. It is reasonable to say that the result converged when the grid size is smaller than 5m.

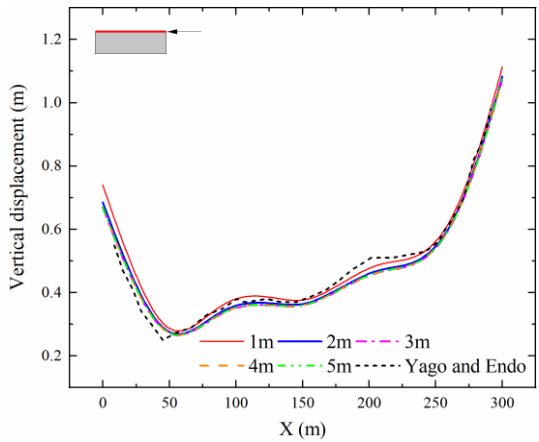

(a) 180m 180degree-S

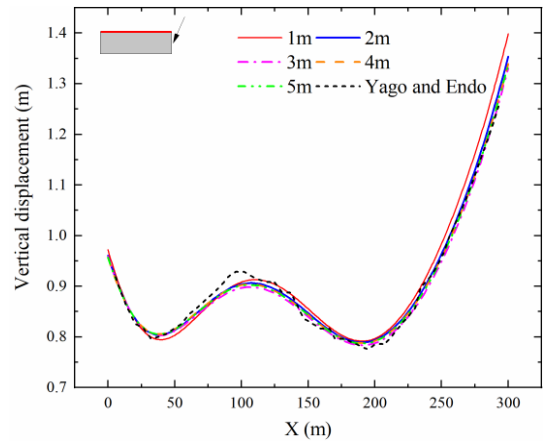

(g) 180m 240degree-S

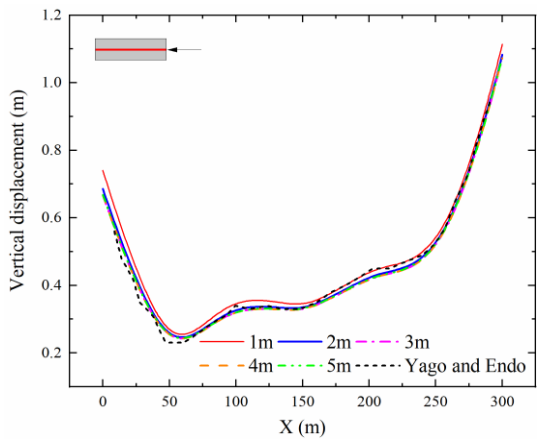

(b) 180m 180degree-C

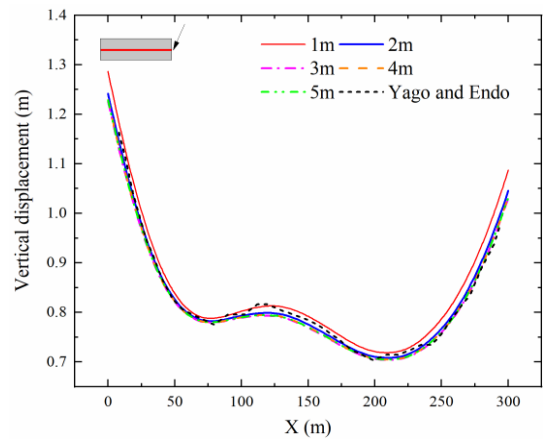

(h) 180m 240degree-C

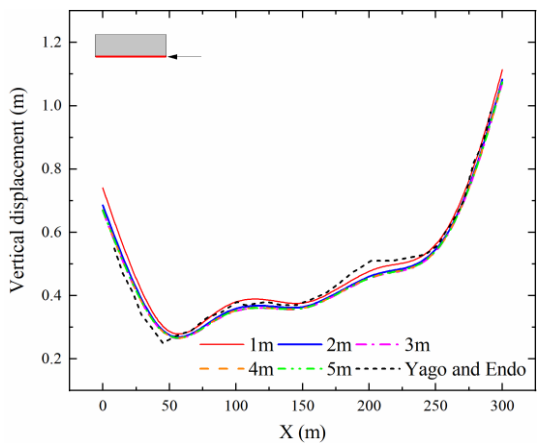

(c) 180m 180degree-P

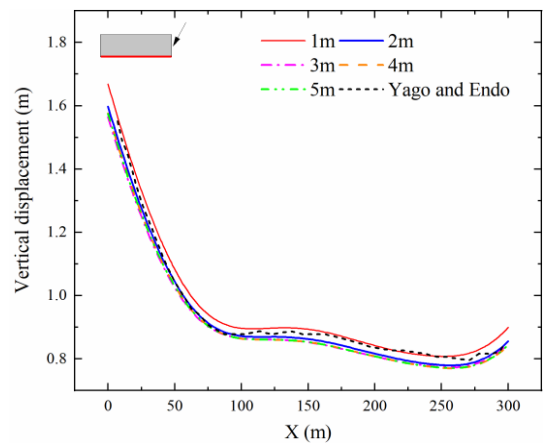

(i) 180m 240degree-P

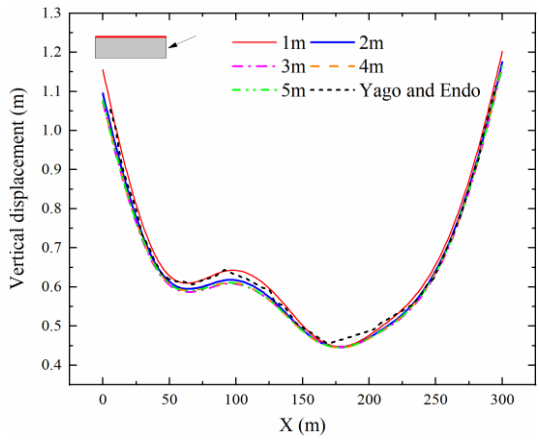
(d) 180m 210degree-S

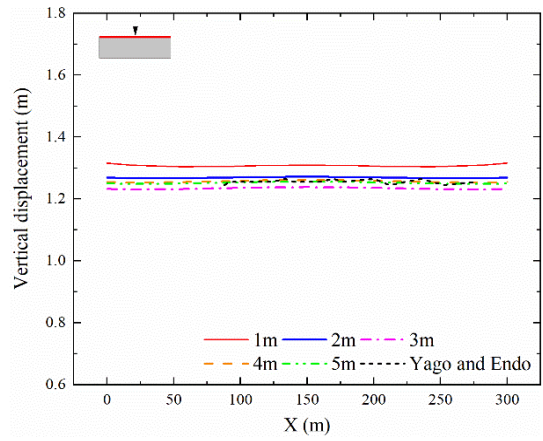
(j) 180m 270degree-S

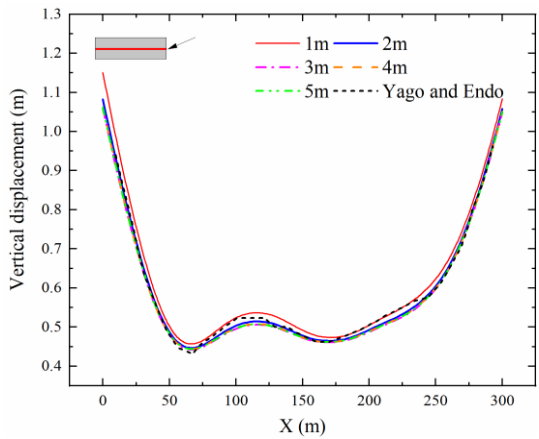
(e) 180m 210degree-C

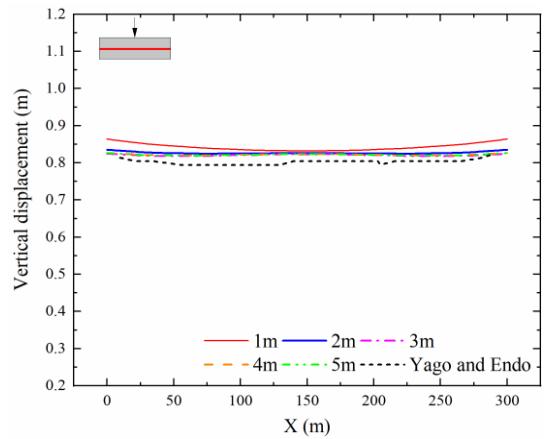
(k) 180m 270degree-C

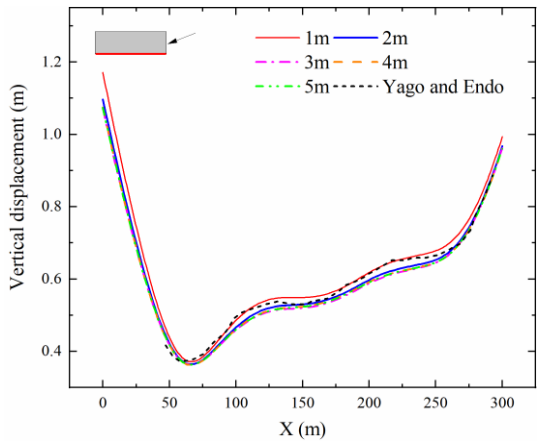
(f) 180m 210degree-P

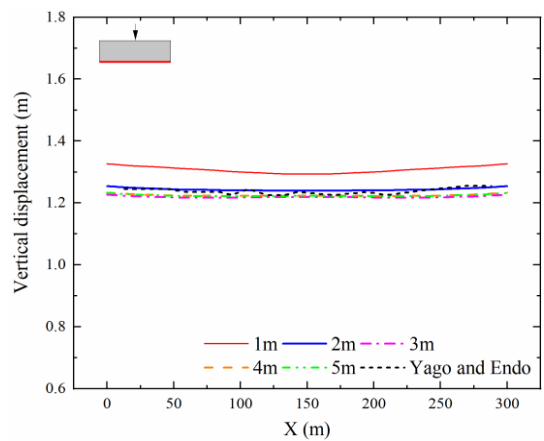
(l) 180m 270degree-P

**Fig. 18. Vertical displacement of MF-300 under a 180m regular wave for four incident angles (180°, 210°, 240° and 270°). Simulations of Yago and Endo [19] using the mode-superposition method is shown in black dashed lines as comparison. Numbers in the legend reflect the grid size (see Table 3). A grey rectangular that represents MF-300 is located at the up-left corner of each sub-figure with the red solid line indicating the position and the arrow indicating the wave incident angle.**

### 4.1.3 Hydroelasticity of MF-300

Results given in this section come from the discretization strategy with an $8 \times 3$ macro-submodule division (Fig. 17a) and a grid size of 1m (Fig. 17b). Distributions of displacement response amplitude of MF-300 under a 180m regular wave are given in Fig. 19. Symmetric features are seen in heading wave (Fig. 19a) and beam wave cases (Fig. 19d).

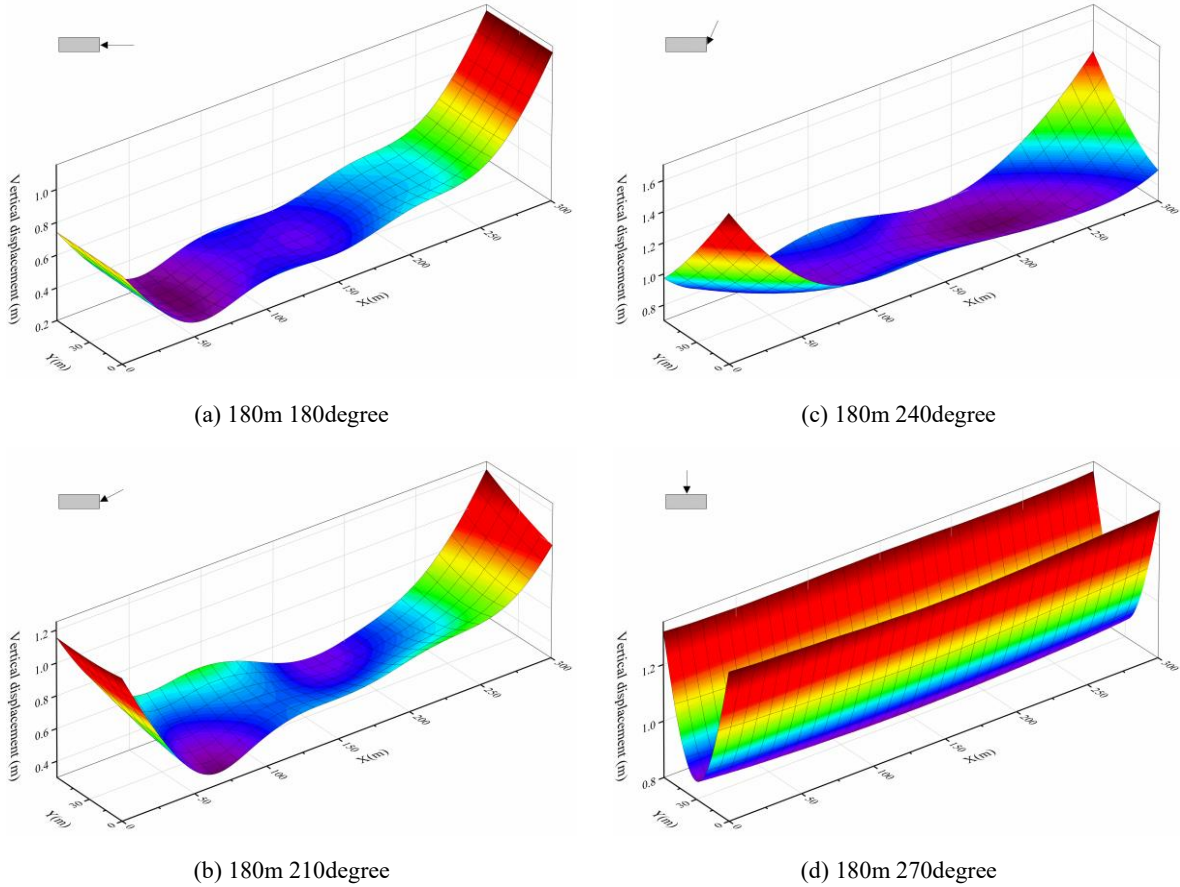

(a) 180m 180degree  (c) 180m 240degree

(b) 180m 210degree  (d) 180m 270degree

**Fig. 19. Vertical displacement distribution of MF-300 under a 180m regular wave. The results come from the discretization strategy with an $8 \times 3$ macro-submodule division (Fig. 17a) and a grid size of 1m (Fig. 17b). The value increases from purple to red, similarly hereinafter.**

Though MF-300 is widely and deeply explored, the authors would like to humbly point out that no result is found in the literature on the bending moment distribution of MF-300 under oblique or beam waves. Only bending moment distribution under heading waves is given. Fig. 20 presents the comparisons between the present results using the spline method and the simulations from Fu et al. [22] on the bending moment distribution $M_y(x,y)\big|_{y=30}$ of MF-300 on the centerline under a 144m regular heading wave (180°). Good agreement is shown, which validates the proposed model.

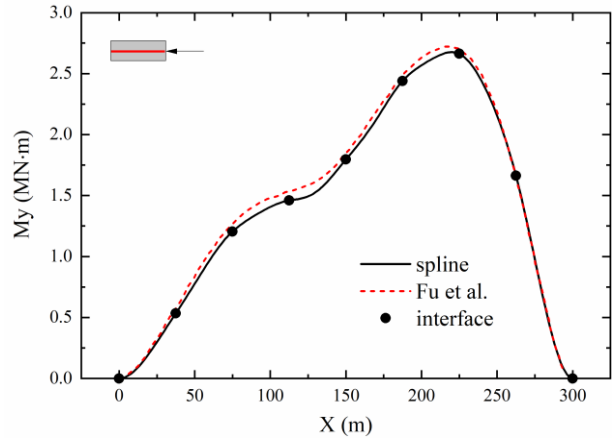

**Fig. 20.** Comparisons on the bending moment distribution $M_y(x,y)\big|_{y=30}$ of MF-300 on the centerline under a 144m regular heading wave between the spline method and Fu et al. [22]. The black circles are the bending moment values on the interfaces. A total of nine ones are shown as the structure is divided into 8 macro-submodules longitudinally (see Fig. 17a).

Bending moment distributions of MF-300 under a 180m regular wave is given in Fig. 21 (results on three specific locations P ($M_y(x,y)\big|_{y=0}$), C ($M_y(x,y)\big|_{y=30}$) and S ($M_y(x,y)\big|_{y=60}$) indicated in Fig. 14) and Fig. 22 (results along the whole structure $M_y(x,y)$). Bending moment $M_y$ in beam wave case is much smaller compared with other cases since the structure barely deforms along the $X$ direction in beam wave case (see Fig. 19d).

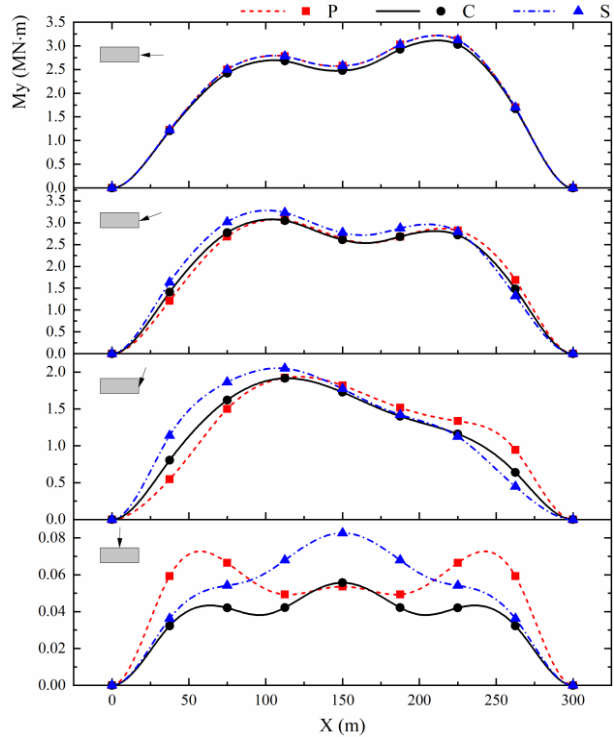

**Fig. 21.** Bending moment distributions on P, C and S of MF-300 under a 180m regular wave for four incident angles. Presented here are results from the spline method.

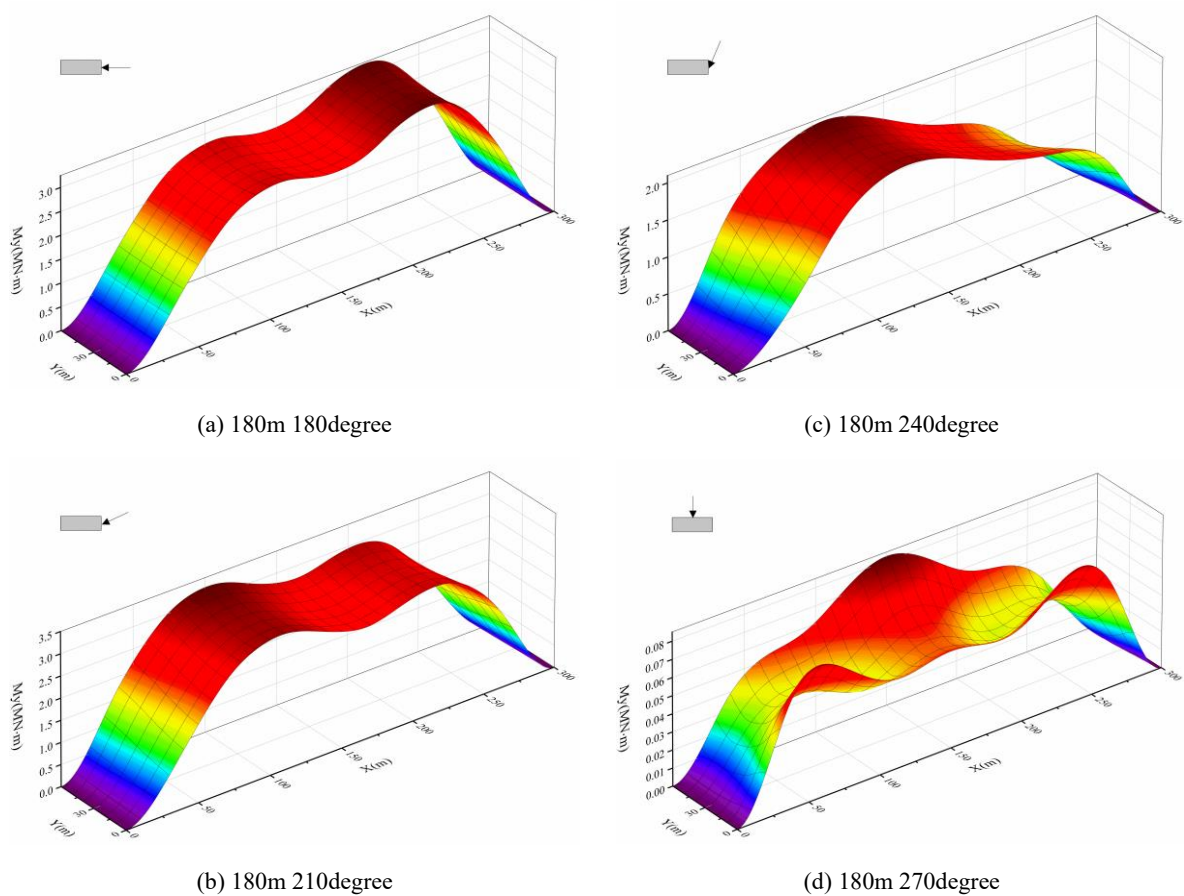

Fig. 22. Bending moment distribution $M_y(x,y)$ along MF-300 under a 180m regular wave for four incident angles. Presented are the results from the spline method.

## 4.2 A square VLFS

Little research is found in the literature on mode-superposition method to deal with a square VLFS. Maybe it is a bit difficult to analyze the hydroelasticity of a square VLFS for the difficulties in modal analysis to finalize the optimal mode combination. Yoon et al. [23] adopted the direct method to calculate the hydroelasticity of a square VLFS under a 180m regular wave, whose characteristics are shown in Table 4. The DMFE method divides the structure into 64 macro submodules with 8 macro-submodules in both $X$ and $Y$ direction (Fig. 23a). Each macro-submodule is a $37.5\text{m} \times 37.5\text{m} \times 4\text{m}$ cuboid as uniform division is taken. Detailed finite element discretization is shown in Fig. 23b with a grid size of 2m taken.

Yoon et al. [23] only gives the hydroelasticity of this square VLFS under a 180m regular wave with a 225° incidence angle. Comparisons are given as follows.

Table 4. Principal particulars of the square VLFS studied by Yoon et al. [23] (Prototype)

| | |
|---|---|
| Length (m) | 300.0 |
| Breadth (m) | 300.0 |
| Draft (m) | 1.1 |
| Depth (m) | 4.0 |
| Vertical bending stiffness (N·m$^2$) | 4.53e8 |
| Water depth (m) | Inf |
| Mass (kg) | 1.01475e8 |

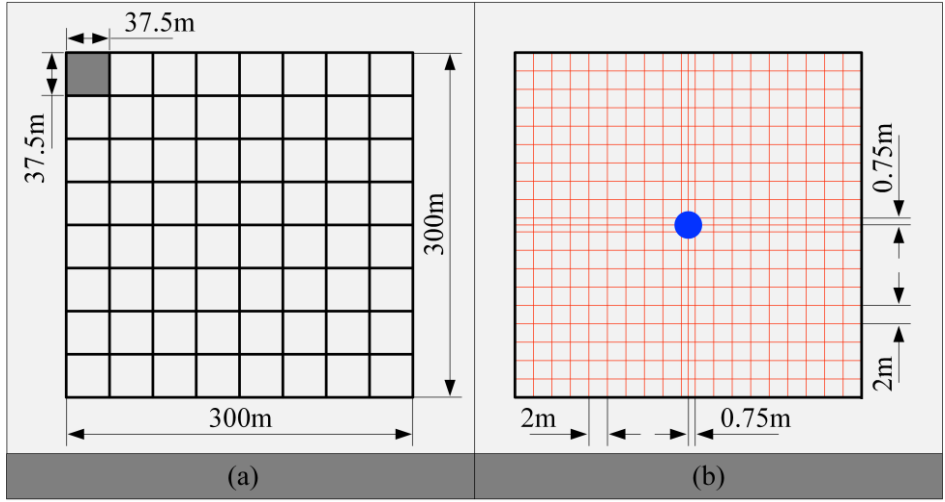

**Fig. 23.** Macro-submodule division (a) and finite element discretization (b) of the square VLFS studied in [23]. The structure is uniformly divided into 8 macro-submodules in both $X$ and $Y$ direction. A grid size of 2m is taken and adjustment is made on elements near the lumped mass.

### 4.2.1 Displacement response

Fig. 24 presents the comparisons on the displacement response on three specific locations between the DMFE method and simulations from Yoon et al. [23]. Rather good agreement is shown. Vertical displacement response along the whole structure is given in Fig. 25, and a diagonally symmetric feature is shown.

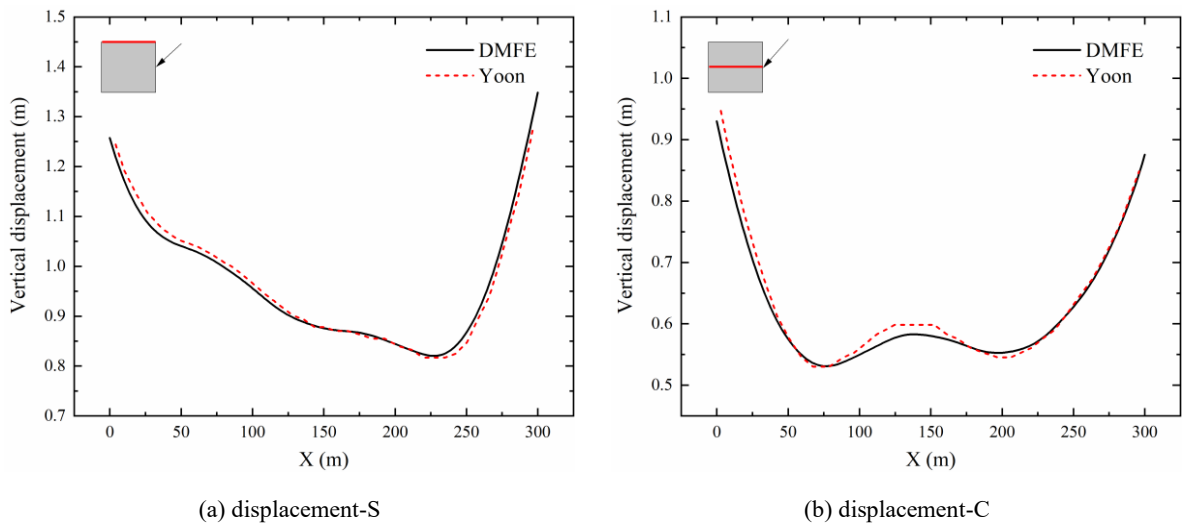

(a) displacement-S

(b) displacement-C

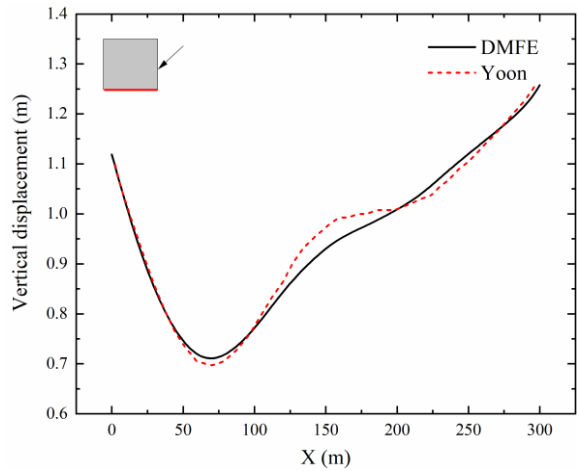

(c) displacement-P

**Fig. 24. Vertical displacement of the square VLFS studied in [23] under a 180m regular wave with 225° incidence angle. A grey square is located at the up-left corner of each sub-figure with the red solid line indicating the position and the arrow indicating the wave incident angle.**

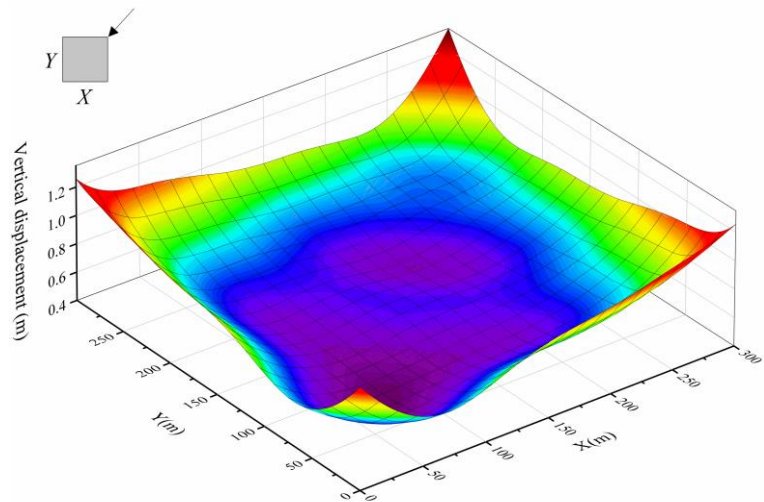

**Fig. 25. Vertical displacement distribution given by the DMFE method of the square VLFS investigated in [23].**

### 4.2.2 Bending moment distribution

Fig. 26 presents the bending moment distributions of the square VLFS studied by Yoon et al. [23] on three locations, i.e., P ($M_y(x,y)\big|_{y=0}$), C ($M_y(x,y)\big|_{y=150}$) and S ($M_y(x,y)\big|_{y=300}$).

We would like to first humbly point out, without any disrespect to Yoon et al., two unreasonable aspects of Yoon et al.'s distribution curve. The first one is non-zero values at free ends. Internal bending moment should be zero at free ends, while a non-ignorable value is observed in Yoon et al. The second one is abrupt changes in distribution curve (Figs. 26a, c). Since the hydrodynamic load is distributed and the structure itself remains homogenous, continuous, uniform and linear, the bending moment distribution is supposed to be smooth, while some unexpected abrupt changes are seen in Yoon et al.

The present results given by the DMFE hydroelasticity method together with the spline based bending moment recovery approach display almost identical trends with Yoon et al, despite that the present values are a bit smaller. However, it has been pointed out that non-zero values are observed at free ends in Yoon et al. Besides, the present results are smooth with no abrupt changes, which seems to be a bit more convincible.

Bending moment distribution $M_y(x,y)$ of this square VLFS under a 180m regular wave is given in Fig. 27.

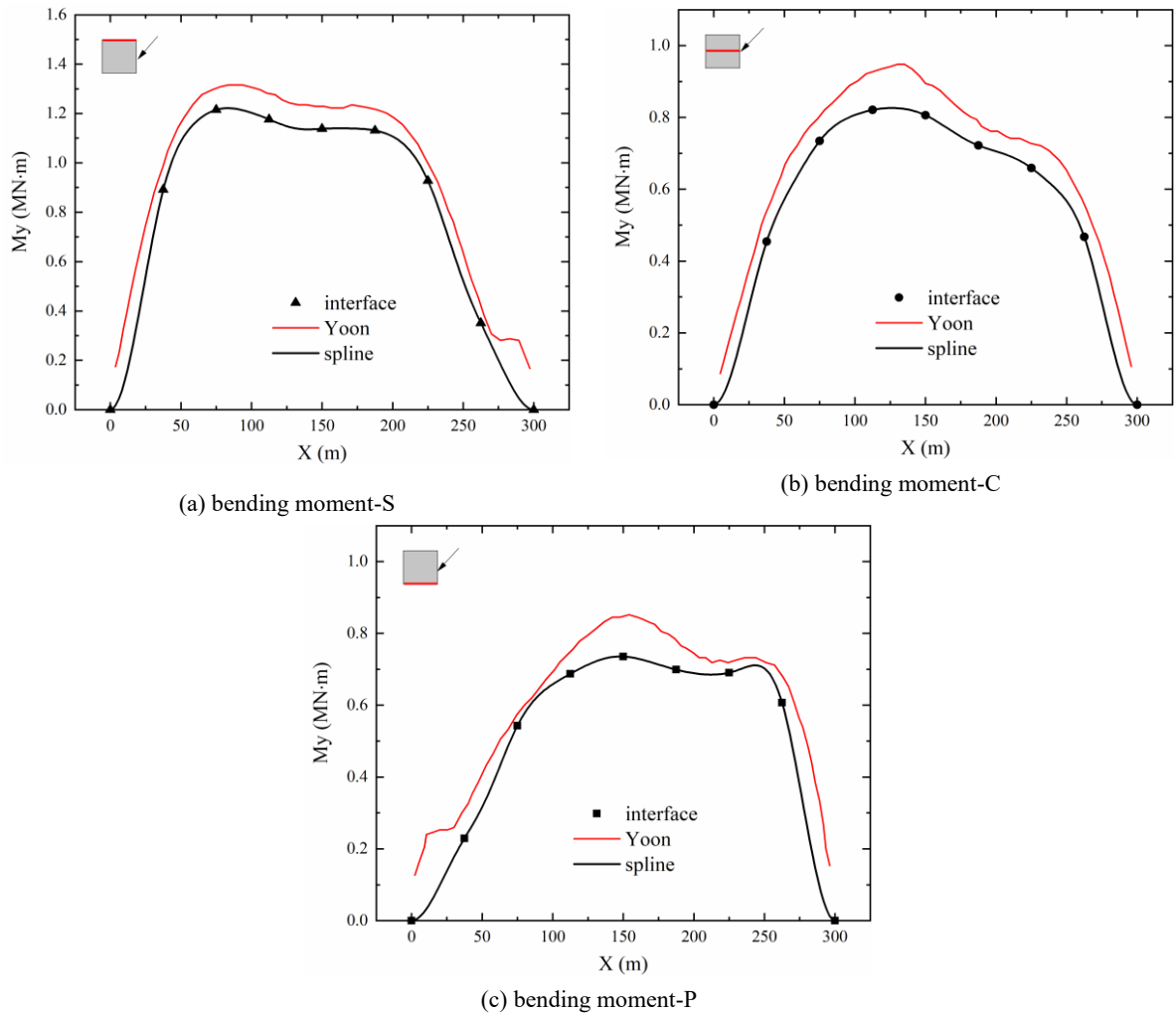

(a) bending moment-S

(b) bending moment-C

(c) bending moment-P

**Fig. 26. Bending moment distribution of the square VLFS studied in [23] on three locations, that is, P, C and S, under a 180m regular wave with a 225° incidence angle.**

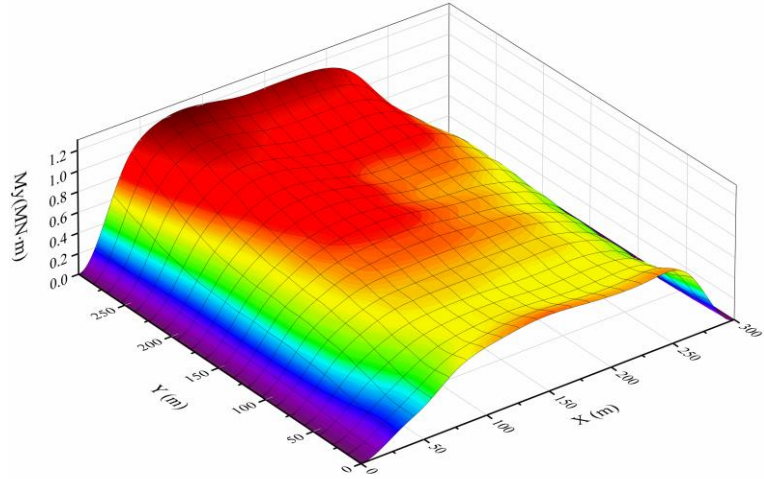

**Fig. 27. Bending moment distribution $M_y(x, y)$ of the square VLFS in [23] under a 180m regular wave with a 225-degree incident angle. The spline method is used.**

All results on the continuous VLFS in Yoon et al. [23] have been re-presented and compared using the proposed DMFE method by far. It is emphasized again that little result is found towards hydroelasticity of a square VLFS, while hydroelasticity of a narrow VLFS is widely and thoroughly studied. Given this, the authors would like to present more results on the hydroelasticity of a square VLFS that may be useful as database for future research. The square VLFS is named as 'SV-300' (**S**quare **V**LFS with a **300**m length and **300**m width) with parameters shown in Table 5 and additional results presented in **Appendix D**.

**Table 5. Principal particulars of 'SV-300' (Prototype)**

| | |
|---|---|
| Length (m) | 300.0 |
| Breadth (m) | 300.0 |
| Draft (m) | 0.5 |
| Depth (m) | 2.0 |
| Vertical bending stiffness (N·m$^2$) | 4.77e11 |
| Water depth (m) | 58.5 |

# 5 Further thoughts

From the comparisons on the displacement and bending moment response with other scholars, it is reasonable to say the DMFE method successfully solves the hydroelasticity of flexible floating structure with a comparable length/width ratio. Although the DMFE method is only implemented on cuboid VLFSs in this paper, the authors are confident that the DMFE method is able to deal with the hydroelasticity of flexible floating structure of arbitrary shape.

However, it is emphasized that the flexible floating structure studied in this paper is a regular cuboid, in which case, the spline method is operational. Fig. 28 gives some conceptual civil designs of large flexible floating structures with rather complicated shape. It is nearly impossible to ensure that a certain number of interfaces distribute uniformly on a given location. Therefore, a least square method is developed and elaborated as follows.

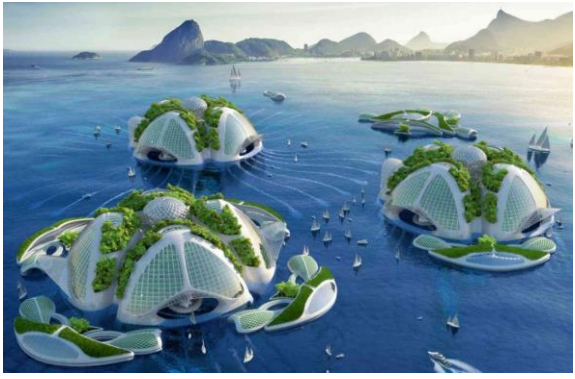 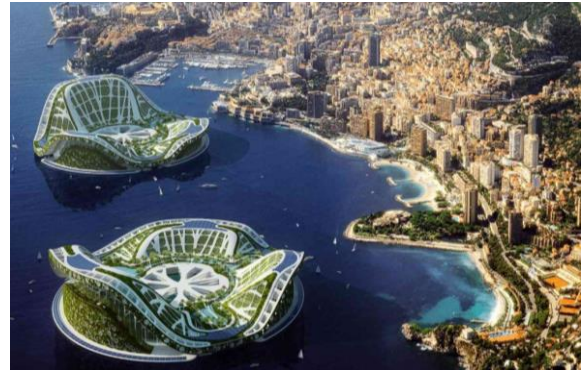

(a) Aequorea project    (b) Lilypad project

**Fig. 28.** Conceptual designs of large flexible floating structure, (a) the Aequorea project – oceanscraper; (b) the Lilypad project – floating ecopolis. Source: https://www.vincent.callebaut.org/.

## 5.1 Least square method for recovery of internal force/moment

Derivation of $M_y(x,y)$ is shown here for example. Based on analysis in Section 2.5 (Figs. 10b and 12), for each macro-submodule, the shear force value $N_y$ and bending moment value $M_y$ of boundary nodes on edges normal to the $X$ axis are calculated through Eqs. 15 and 18. Therefore, the least square method is performed to calculate the shear force or bending moment at a given location of each macro-submodule using the already known values on the boundary of the macro-submodule as a basis. Denote the bending moment distribution of macro-submodule $(m,n)$, $M_y^{(m,n)}(x,y)$ as a polynomial function up to the third order,

$$M_y^{(m,n)}(x,y) = \sum_{j=1}^{10} a_j \varphi_j(x,y) \\ = a_1 + a_2 x + a_3 y + a_4 x^2 + a_5 xy + a_6 y^2 + a_7 x^3 + a_8 x^2 y + a_9 xy^2 + a_{10} y^3 \tag{23}$$

where $a_1, a_2, \cdots, a_{10}$ are the undetermined coefficients.

Then, the shear force distribution is expressed as

$$N_y^{(m,n)}(x,y) = \frac{\partial}{\partial x} M_y^{(m,n)}(x,y) = \sum_{j=1}^{10} a_j \frac{\partial}{\partial x} \varphi_j(x,y) \\ = a_2 + 2a_4 x + a_5 y + 3a_7 x^2 + 2a_8 xy + a_9 y^2 \tag{24}$$

The authors put a great thought into the order of the polynomial function. First, the order should be at least three to ensure the non-linear features in bending moment and shear force distribution. Second, the proposed least square method in the frame work of the DMFE method should cover the bending moment recovery in the DMB method where a third-order polynomial function with respect to only $x$ is used. Therefore, the bending moment distribution should degenerate into that function when the variable $y$ is fixed at a chosen value. Last, high order interpolation may product Runge phenomenon and the number of the undetermined coefficients will increase. After some trials, the cubic polynomial function is adopted.

Detailed derivation of $a_1, a_2, \cdots, a_{10}$ is presented in **Appendix E**. Repetition of the method on every macro-submodule gives the bending moment distribution $M_y(x,y)$ along the structure. Identical strategies and procedures are used to calculate $M_x(x,y)$.

## 5.2 The black cloud

It can be seen that the least square method is performed on each macro-submodule regardless of its shape. With known shear force and bending moment at specific boundary nodes, an analytical formula will be derived to describe the bending moment distribution surface on this macro-submodule. Theoretically speaking, the least square method is better than the spline method for the former considers not only the bending moment but the shear force at boundary nodes. However, there is one deficiency of the least square method that the authors cannot conquer, and that is why the least square method is given at last even though the spline method is accurate enough to recover bending moment distribution of a square VLFS.

As shown in Fig. 23, the square VLFS is divided into 8 macro-submodules in both $X$ and $Y$ direction. When calculating the bending moment distribution $M_y(x,y)$, shear force and bending moment values on the nine blue lines, which are indicated in Fig. 29, are calculated. Fig. 30 gives the internal force distribution on these blue lines with (a) the shear force and (b) the bending moment. Notice $X = 0$ and $X = 300$ are with zero internal force for being free ends. It can be seen that the bending moment distributions change gradually along the $Y$ direction. However, it is rather strange to see that the shear force distribution increase abruptly near the free ends. Therefore, the least square method could be performed taking account of all nodes and performed eliminating the nodes with suspiciously large shear force value. The former is denoted as LS-all (**L**east **S**quare-**all** nodes) and the latter LS-part.

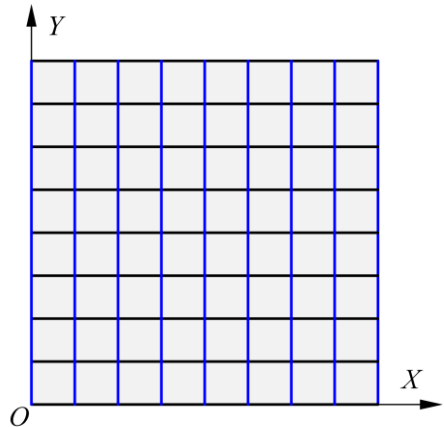

Fig. 29. Shear force $N_y$ and bending moment $M_y$ value on the blue lines are calculated in performing the least square method to recover bending moment distribution $M_y(x,y)$.

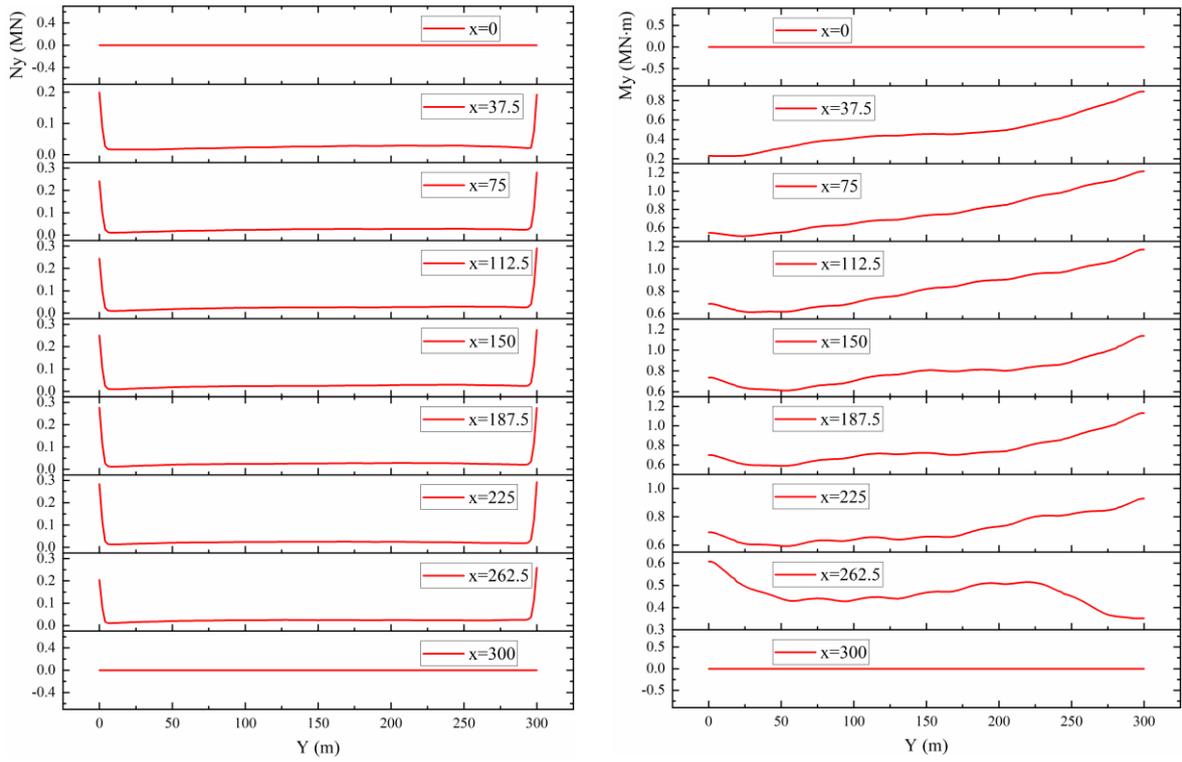

(a) shear force  (b) bending moment

**Fig. 30.** Shear force $N_y(x,y)\big|_{x=x_0}$ and bending moment $M_y(x,y)\big|_{x=x_0}$ distribution on the blue lines indicated in **Fig. 29.**

Bending moment distribution of the square VLFS studied by Yoon et al. [23] under a 180m regular wave with a 225° incidence angle calculated with the least square method is given in Fig. 31 (on P, C and S) and Fig. 32 (along the whole structure), which can be compared with the spline-method result in Fig. 27. Large oscillating phenomena are seen in the LS-all results near the free ends in the $Y$ direction, resulting from the suspiciously large shear force values (large slope) at these locations. Slight difference is observed between the LS-part results and the spline results, which somehow indicates the accuracy of the least square method. Bending moment distribution on C for LS-all and LS-part coincides since the abrupt increase of shear force imposes no effect on the bending moment distribution of macro-submodules without nodes near the free ends. Note that the least square method gives an optimal solution for each macro-submodule while the spline method gives an optimal solution for a given location. Therefore, when the LS results are combined together to form the bending moment distribution curve on C, it diverges from the spline result.

However, both the bending moment and the shear force are calculated through Eq. 15. The spline method uses the bending moment only and gives a satisfactory result. It is rather strange to see the shear force displays an incomprehensible steep increase towards the free ends. The authors have spent an amount of effort but still cannot conquer this.

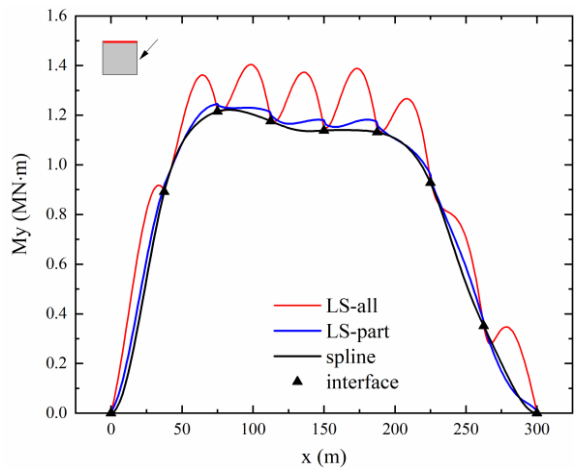

(a) bending moment-S

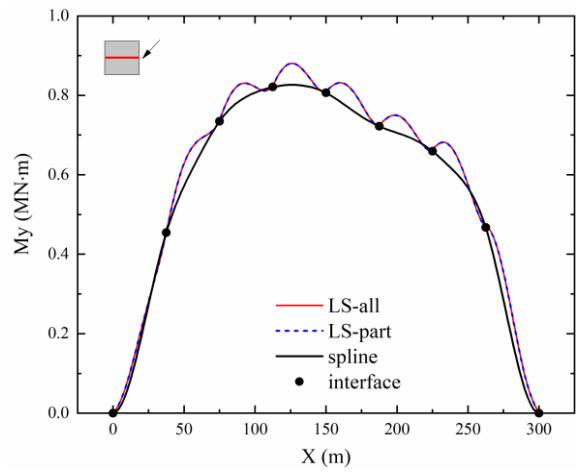

(b) bending moment-C

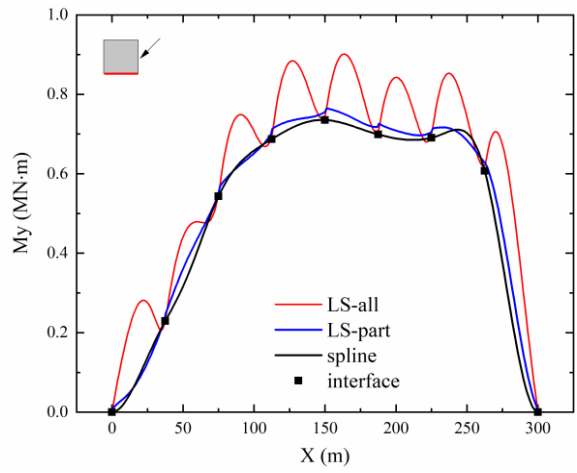

(c) bending moment-P

**Fig. 31. Comparisons of bending moment distribution on three locations P, C and S, where LS-all refers to the least square method taking account of every node, LS-part refers to the least square method eliminating nodes with suspiciously large shear force value, spline refers to the spline method.**

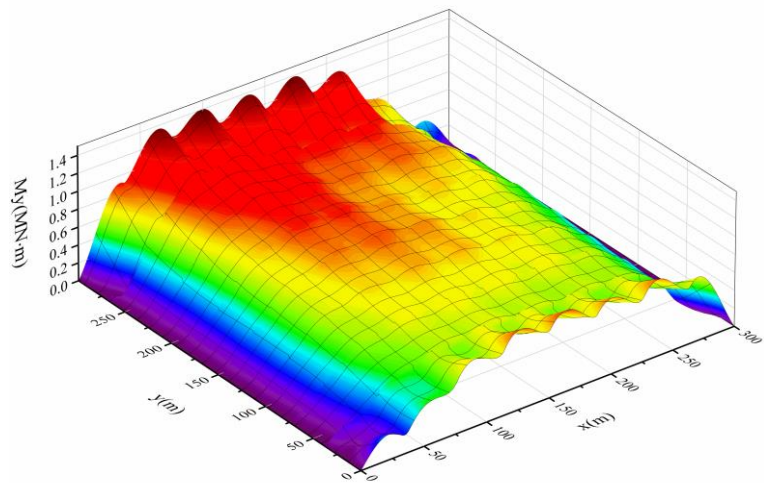

(a) Least square-all

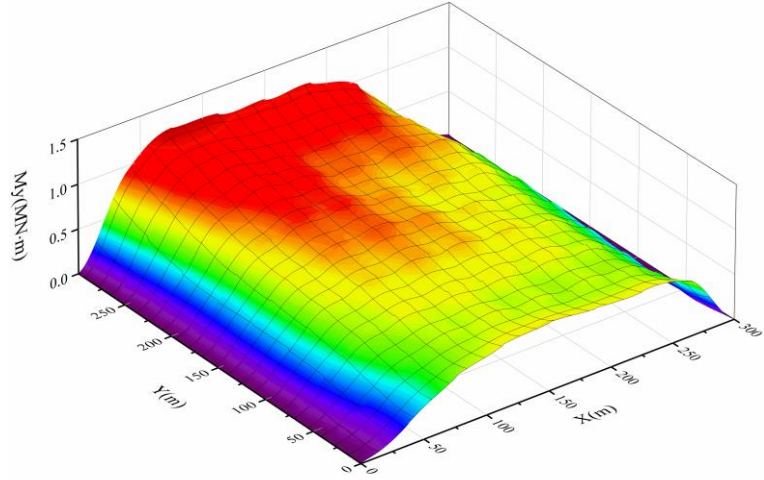

(b) Least square-part

**Fig. 32. Bending moment distribution $M_y(x,y)$ of the square VLFS investigated by Yoon et al. under a 180m regular wave with a 45-degree incident angle.**

# 6 Conclusion

A discrete-module-finite-element method (DMFE) is developed in the frequency domain to analyze hydroelasticity of large flexible floating structures, including the displacement and internal force responses.

In the framework of DMFE method, a large floating flexible structure is first discretized in 'two dimensions' (i.e., along the length and width direction of the structure), resulting in freely floating macro-submodules. A multi-rigid-body hydrodynamic analysis based on linear potential flow theory is performed to obtain the external forces on these macro-submodules, including the wave excitation force, radiation force, hydrostatic force and inertial force. Subsequently, all submodules are abstracted to be lumped masses (at centers of gravity of the submodules) that subject to the total external forces (with displacements as unknown variables).

Apart from the external forces, these lumped masses are also subjected to structural deformation induced forces due to the flexibility of the original structure. The structural force on all lumped masses can be derived based on the overall stiffness matrix with respect to all lumped masses and the unknown displacement variables of these lumped masses. A finite element discretization is introduced to the floating flexible structures to obtain the stiffness matrix for all submodules. After some deliberate substructuring and matrix manipulations, the lumped-mass stiffness matrix with respect to all lumped masses can be derived.

Finally, the hydroelastic equation of the floating flexible structure is established (at positions of all lumped masses, with displacements as unknown variables) according to the force equilibrium condition between the external and structural forces following the d'Alembert principle. The displacements of lumped masses are obtained by solving this hydroelastic equation.

Recovery of the displacement response at any given position of the flexible structure is reverse to the derivation of the lumped-mass stiffness matrix. Recovery of the internal force of the structure is implemented with a spline interpolation using values of bending moment on interfaces of different macro-submodules. The DMFE method is applied on a narrow very large floating structure (VLFS) and a square VLFS. Good agreement is shown in comparisons with other scholars.

Besides, the authors describe a least square method to recover bending moment distribution of arbitrary-shaped flexible floating structure, but with one unsolved problem that the shear force increases abruptly near free ends. The authors are open to all useful opinions. Despite this black cloud, the DMFE method developed in Section 2 is clearly qualified to deal with the hydroelasticity of square VLFSs.

The DMFE method takes a huge step forward from the discrete-module-beam (DMB) method by implementing the finite element to deal with hydroelasticity of flexible floating structures with a comparable length/width ratio. Compared with the mode-superposition method, the DMFE method needs no modal analysis to finalize the optimal mode combination that may be uneasy for a large floating flexible structure with a comparable length/width ratio or with a complicated shape. Compared with the direct method, the DMFE method adopts the finite element but reduces the matrix size greatly by substructuring and matrix manipulations, which improves the computational efficiency.

Proposition of the DMFE method open up a number possibilities for future work. First, the DMFE method could be verified on a VLFS with complicated shape. Then, hydroelastic experiments on a square VLFS could be done to fill the void in the literature. In addition, it is worth exploring the following aspects, such as, modelling of hinge connection, extension into the time domain, strategies to deal with concentrated force, hydroelasticity under unsteady external loads, consideration of mooring lines, non-linear properties of both the hydrodynamic loads and the structure and so forth.

# Acknowledgement

This work is supported by Hainan Provincial Natural Science Foundation of China (Grant no. 520QN290), State Key Laboratory of Coastal and Offshore Engineering (Grant No. LP2019) and the 2020 Research Program of Sanya Yazhou Bay Science and Technology City (Grant No. SKJC-2020-01-006).

# Appendix A

The hydrodynamic coefficients, added mass $\mathbf{A}(\omega)$, radiation damping $\mathbf{B}(\omega)$, hydrostatic restoring force $\mathbf{C}$ and wave excitation force $\mathbf{F}_E$, are defined as follows:

$$\mathbf{A}(\omega) = \begin{bmatrix} \mathbf{A}^{[(1,1),(1,1)]} & \mathbf{A}^{[(1,1),(1,2)]} & \cdots & \mathbf{A}^{[(1,1),(M,N)]} \\ \mathbf{A}^{[(2,2),(1,1)]} & \mathbf{A}^{[(2,2),(2,2)]} & \cdots & \mathbf{A}^{[(2,2),(M,N)]} \\ \vdots & \vdots & \ddots & \vdots \\ \mathbf{A}^{[(M,N),(1,1)]} & \mathbf{A}^{[(M,N),(2,2)]} & \cdots & \mathbf{A}^{[(M,N),(M,N)]} \end{bmatrix}_{6MN \times 6MN} \quad (A1)$$

where $\mathbf{A}^{[(m,n),(p,q)]}$ is a $6 \times 6$ matrix that represents the added mass coefficients matrix of the macro-submodule $(m,n)$ induced by the motion of the macro-submodule $(p,q)$.

$$\mathbf{B}(\omega) = \begin{bmatrix} \mathbf{B}^{[(1,1),(1,1)]} & \mathbf{B}^{[(1,1),(1,2)]} & \cdots & \mathbf{B}^{[(1,1),(M,N)]} \\ \mathbf{B}^{[(2,2),(1,1)]} & \mathbf{B}^{[(2,2),(2,2)]} & \cdots & \mathbf{B}^{[(2,2),(M,N)]} \\ \vdots & \vdots & \ddots & \vdots \\ \mathbf{B}^{[(M,N),(1,1)]} & \mathbf{B}^{[(M,N),(2,2)]} & \cdots & \mathbf{B}^{[(M,N),(M,N)]} \end{bmatrix}_{6MN \times 6MN} \quad (A2)$$

where $\mathbf{B}^{[(m,n),(p,q)]}$ is a $6 \times 6$ matrix that represents the radiation damping coefficients matrix of the macro-submodule $(m, n)$ induced by the motion of the macro-submodule $(p, q)$.

$$\mathbf{C} = \begin{bmatrix} \mathbf{C}^{[(1,1),(1,1)]} & 0 & \cdots & 0 \\ 0 & \mathbf{C}^{[(1,2),(1,2)]} & \cdots & 0 \\ \vdots & \vdots & \ddots & \vdots \\ 0 & 0 & \cdots & \mathbf{C}^{[(M,N),(M,N)]} \end{bmatrix}_{6MN \times 6MN} \quad (A3)$$

where $\mathbf{C}^{[(m,n),(m,n)]}$ is a $6 \times 6$ matrix that represents the hydrostatic restoring coefficients matrix of the macro-submodule $(m, n)$.

$$\mathbf{F}_E = \begin{bmatrix} \mathbf{F}_E^{(1,1)} \\ \mathbf{F}_E^{(1,2)} \\ \vdots \\ \mathbf{F}_E^{(M,N)} \end{bmatrix}_{6MN \times 1} \quad (A4)$$

where $\mathbf{F}_E^{(m,n)}$ is a $6 \times 1$ matrix that represents the wave excitation force exerted on the macro-submodule $(m, n)$.

The mass matrix of all macro-submodules $\mathbf{M}$ is expressed as

$$\mathbf{M} = \begin{bmatrix} \mathbf{M}^{[(1,1),(1,1)]} & 0 & \cdots & 0 \\ 0 & \mathbf{M}^{[(1,2),(1,2)]} & \cdots & 0 \\ \vdots & \vdots & \ddots & \vdots \\ 0 & 0 & \cdots & \mathbf{M}^{[(M,N),(M,N)]} \end{bmatrix}_{6MN \times 6MN} \quad (A5)$$

where $\mathbf{M}^{[(m,n),(m,n)]}$ is a $6 \times 6$ matrix that represents the mass matrix of the macro-submodule $(m, n)$ (see [24]). The displacement of all lumped masses $\boldsymbol{\xi}$ is expressed as

$$\boldsymbol{\xi} = \begin{bmatrix} \boldsymbol{\xi}^{(1,1)} \\ \boldsymbol{\xi}^{(1,2)} \\ \vdots \\ \boldsymbol{\xi}^{(M,N)} \end{bmatrix}_{6MN \times 1} \quad (A6)$$

# Appendix B

Detailed form of $\mathbf{k}_{\text{module}}$ is given for better illustration

$$\mathbf{k}_{\text{module}} = \begin{bmatrix} \mathbf{k}_{\text{module}}^{\text{lumped-mass}} \\ \mathbf{k}_{\text{module}}^{\text{boundary}} \\ \mathbf{k}_{\text{module}}^{\text{inner}} \end{bmatrix} = \begin{bmatrix} \mathbf{k}_{\text{lumped-mass}}^{\text{lumped-mass}} & \mathbf{k}_{\text{boundary}}^{\text{lumped-mass}} & \mathbf{k}_{\text{inner}}^{\text{lumped-mass}} \\ \mathbf{k}_{\text{lumped-mass}}^{\text{boundary}} & \mathbf{k}_{\text{boundary}}^{\text{boundary}} & \mathbf{k}_{\text{inner}}^{\text{boundary}} \\ \mathbf{k}_{\text{lumped-mass}}^{\text{inner}} & \mathbf{k}_{\text{boundary}}^{\text{inner}} & \mathbf{k}_{\text{inner}}^{\text{inner}} \end{bmatrix} \quad (B1)$$

Similarities are observed among parameters in Eq. B1 and explanations are made only on the second row for the sake of brevity. $\mathbf{k}_{\text{module}}^{\text{boundary}}$ is the rows of $\mathbf{k}_{\text{module}}$ that corresponds to the boundary nodes with

$$\mathbf{k}_{\text{module}}^{\text{boundary}} \begin{bmatrix} \boldsymbol{\xi}_{\text{lumped-mass}}^{(m,n)} \\ \boldsymbol{\xi}_{\text{boundary}}^{(m,n)} \\ \boldsymbol{\xi}_{\text{inner}}^{(m,n)} \end{bmatrix}$$

$$= \begin{bmatrix} \mathbf{k}_{\text{lumped-mass}}^{\text{boundary}} & \mathbf{k}_{\text{boundary}}^{\text{boundary}} & \mathbf{k}_{\text{inner}}^{\text{boundary}} \end{bmatrix} \begin{bmatrix} \boldsymbol{\xi}_{\text{lumped-mass}}^{(m,n)} \\ \boldsymbol{\xi}_{\text{boundary}}^{(m,n)} \\ \boldsymbol{\xi}_{\text{inner}}^{(m,n)} \end{bmatrix} \quad (B2)$$

$$= \mathbf{k}_{\text{lumped-mass}}^{\text{boundary}} \boldsymbol{\xi}_{\text{lumped-mass}}^{(m,n)} + \mathbf{k}_{\text{boundary}}^{\text{boundary}} \boldsymbol{\xi}_{\text{boundary}}^{(m,n)} + \mathbf{k}_{\text{inner}}^{\text{boundary}} \boldsymbol{\xi}_{\text{inner}}^{(m,n)}$$

$$= \mathbf{F}_{\text{internal}}^{(m,n)}$$

where $\mathbf{k}_{\text{lumped-mass}}^{\text{boundary}}$ is the columns of $\mathbf{k}_{\text{module}}^{\text{boundary}}$ that corresponds to the lumped mass, $\mathbf{k}_{\text{boundary}}^{\text{boundary}}$ corresponds to the boundary nodes and $\mathbf{k}_{\text{inner}}^{\text{boundary}}$ the inner nodes.

Eq. 13 and Eq. B1 give

$$\begin{bmatrix} \mathbf{k}_{\text{lumped-mass}}^{\text{lumped-mass}} & \mathbf{k}_{\text{boundary}}^{\text{lumped-mass}} & \mathbf{k}_{\text{inner}}^{\text{lumped-mass}} \\ \mathbf{k}_{\text{lumped-mass}}^{\text{boundary}} & \mathbf{k}_{\text{boundary}}^{\text{boundary}} & \mathbf{k}_{\text{inner}}^{\text{boundary}} \\ \mathbf{k}_{\text{lumped-mass}}^{\text{inner}} & \mathbf{k}_{\text{boundary}}^{\text{inner}} & \mathbf{k}_{\text{inner}}^{\text{inner}} \end{bmatrix} \begin{bmatrix} \boldsymbol{\xi}_{\text{lumped-mass}}^{(m,n)} \\ \boldsymbol{\xi}_{\text{boundary}}^{(m,n)} \\ \boldsymbol{\xi}_{\text{inner}}^{(m,n)} \end{bmatrix} = \begin{bmatrix} \mathbf{F}_{\text{EXT}}^{(m,n)} \\ \mathbf{F}_{\text{boundary}}^{(m,n)} \\ \mathbf{0} \end{bmatrix} \quad (B3)$$

Some manipulations give

$$\boldsymbol{\xi}_{\text{inner}}^{(m,n)} = -\left[\mathbf{k}_{\text{inner}}^{\text{inner}}\right]^{-1} \begin{bmatrix} \mathbf{k}_{\text{lumped-mass}}^{\text{inner}} & \mathbf{k}_{\text{boundary}}^{\text{inner}} \end{bmatrix} \begin{bmatrix} \boldsymbol{\xi}_{\text{lumped-mass}}^{(m,n)} \\ \boldsymbol{\xi}_{\text{boundary}}^{(m,n)} \end{bmatrix} \quad (B4)$$

$$\triangleq -\mathbf{\Lambda}_{\text{INNER}} \begin{bmatrix} \boldsymbol{\xi}_{\text{lumped-mass}}^{(m,n)} \\ \boldsymbol{\xi}_{\text{boundary}}^{(m,n)} \end{bmatrix}$$

$$\begin{bmatrix} \mathbf{F}_{\text{EXT}}^{(m,n)} \\ \mathbf{F}_{\text{boundary}}^{(m,n)} \end{bmatrix}$$

$$= \left\{ \begin{bmatrix} \mathbf{k}_{\text{lumped-mass}}^{\text{lumped-mass}} & \mathbf{k}_{\text{boundary}}^{\text{lumped-mass}} \\ \mathbf{k}_{\text{lumped-mass}}^{\text{boundary}} & \mathbf{k}_{\text{boundary}}^{\text{boundary}} \end{bmatrix} - \begin{bmatrix} \mathbf{k}_{\text{inner}}^{\text{lumped-mass}} \\ \mathbf{k}_{\text{inner}}^{\text{boundary}} \end{bmatrix} \mathbf{\Lambda}_{\text{INNER}} \right\} \begin{bmatrix} \boldsymbol{\xi}_{\text{lumped-mass}}^{(m,n)} \\ \boldsymbol{\xi}_{\text{boundary}}^{(m,n)} \end{bmatrix} \quad \text{(B5)}$$

$$\triangleq \mathbf{K}_{\text{OUTER}} \begin{bmatrix} \boldsymbol{\xi}_{\text{lumped-mass}}^{(m,n)} \\ \boldsymbol{\xi}_{\text{boundary}}^{(m,n)} \end{bmatrix}$$

# Appendix C

$\mathbf{K}_{\text{OUTER}}$ is expressed as

$$\mathbf{K}_{\text{OUTER}} = \begin{bmatrix} \mathbf{K}_{\text{OUTER}}^{\text{lumped-mass}} \\ \mathbf{K}_{\text{OUTER}}^{\text{boundary}} \\ \mathbf{K}_{\text{OUTER}}^{\text{free-edge}} \end{bmatrix} = \begin{bmatrix} \mathbf{K}_{\text{lumped-mass}}^{\text{lumped-mass}} & \mathbf{K}_{\text{boundary}}^{\text{lumped-mass}} & \mathbf{K}_{\text{free-edge}}^{\text{lumped-mass}} \\ \mathbf{K}_{\text{lumped-mass}}^{\text{boundary}} & \mathbf{K}_{\text{boundary}}^{\text{boundary}} & \mathbf{K}_{\text{free-edge}}^{\text{boundary}} \\ \mathbf{K}_{\text{lumped-mass}}^{\text{free-edge}} & \mathbf{K}_{\text{boundary}}^{\text{free-edge}} & \mathbf{K}_{\text{free-edge}}^{\text{free-edge}} \end{bmatrix} \quad \text{(C1)}$$

Meanings of each sub-matrix in Eq. C1 could be referred to $\mathbf{k}_{module}$ in Eqs. B1-B2. Apparently,

$$\begin{bmatrix} \mathbf{K}_{\text{lumped-mass}}^{\text{lumped-mass}} & \mathbf{K}_{\text{non-free}}^{\text{lumped-mass}} & \mathbf{K}_{\text{free-edge}}^{\text{lumped-mass}} \\ \mathbf{K}_{\text{lumped-mass}}^{\text{non-free}} & \mathbf{K}_{\text{non-free}}^{\text{non-free}} & \mathbf{K}_{\text{free-edge}}^{\text{non-free}} \\ \mathbf{K}_{\text{lumped-mass}}^{\text{free-edge}} & \mathbf{K}_{\text{non-free}}^{\text{free-edge}} & \mathbf{K}_{\text{free-edge}}^{\text{free-edge}} \end{bmatrix} \begin{bmatrix} \boldsymbol{\xi}_{\text{lumped-mass}}^{(m,n)} \\ \boldsymbol{\xi}_{\text{non-free}}^{(m,n)} \\ \boldsymbol{\xi}_{\text{free-edge}}^{(m,n)} \end{bmatrix} = \begin{bmatrix} \mathbf{F}_{\text{EXT}}^{(m,n)} \\ \mathbf{F}_{\text{non-free}}^{(m,n)} \\ \mathbf{0} \end{bmatrix} \quad \text{(C2)}$$

Some manipulations give

$$\boldsymbol{\xi}_{\text{free-edge}}^{(m,n)} = -\left[\mathbf{K}_{\text{free-edge}}^{\text{free-edge}}\right]^{-1} \begin{bmatrix} \mathbf{K}_{\text{lumped-mass}}^{\text{free-edge}} & \mathbf{K}_{\text{non-free}}^{\text{free-edge}} \end{bmatrix} \begin{bmatrix} \boldsymbol{\xi}_{\text{lumped-mass}}^{(m,n)} \\ \boldsymbol{\xi}_{\text{non-free}}^{(m,n)} \end{bmatrix}$$

$$\triangleq -\mathbf{\Lambda}_{\text{FREE-EDGE}} \begin{bmatrix} \boldsymbol{\xi}_{\text{lumped-mass}}^{(m,n)} \\ \boldsymbol{\xi}_{\text{non-free}}^{(m,n)} \end{bmatrix} \quad \text{(C3)}$$

$$\begin{bmatrix} \mathbf{F}_{\text{EXT}}^{(m,n)} \\ \mathbf{F}_{\text{non-free}}^{(m,n)} \end{bmatrix}$$
$$= \left\{ \begin{bmatrix} \mathbf{K}_{\text{lumped-mass}}^{\text{lumped-mass}} & \mathbf{K}_{\text{non-free}}^{\text{lumped-mass}} \\ \mathbf{K}_{\text{lumped-mass}}^{\text{non-free}} & \mathbf{K}_{\text{non-free}}^{\text{non-free}} \end{bmatrix} - \begin{bmatrix} \mathbf{K}_{\text{free-edge}}^{\text{lumped-mass}} \\ \mathbf{K}_{\text{free-edge}}^{\text{non-free}} \end{bmatrix} \mathbf{\Lambda}_{\text{FREE-EDGE}}^{(m,n)} \right\} \begin{bmatrix} \boldsymbol{\xi}_{\text{lumped-mass}}^{(m,n)} \\ \boldsymbol{\xi}_{\text{non-free}}^{(m,n)} \end{bmatrix} \quad (C4)$$
$$\triangleq \mathbf{K}_{\text{NON-FREE}} \begin{bmatrix} \boldsymbol{\xi}_{\text{lumped-mass}}^{(m,n)} \\ \boldsymbol{\xi}_{\text{boundary}}^{(m,n)} \end{bmatrix}$$

# Appendix D

Displacement responses are given first on SV-300 under a 180m regular wave for five incident angles (180°, 210°, 225°, 240° and 270°). Fig. D1 gives the vertical displacement along 5 specific locations, that is, P, C and S indicated in Fig. 14 and two additional ones, $Y = 75$ and $Y = 225$. Fig. D2 presents vertical displacement distribution along the whole structure.

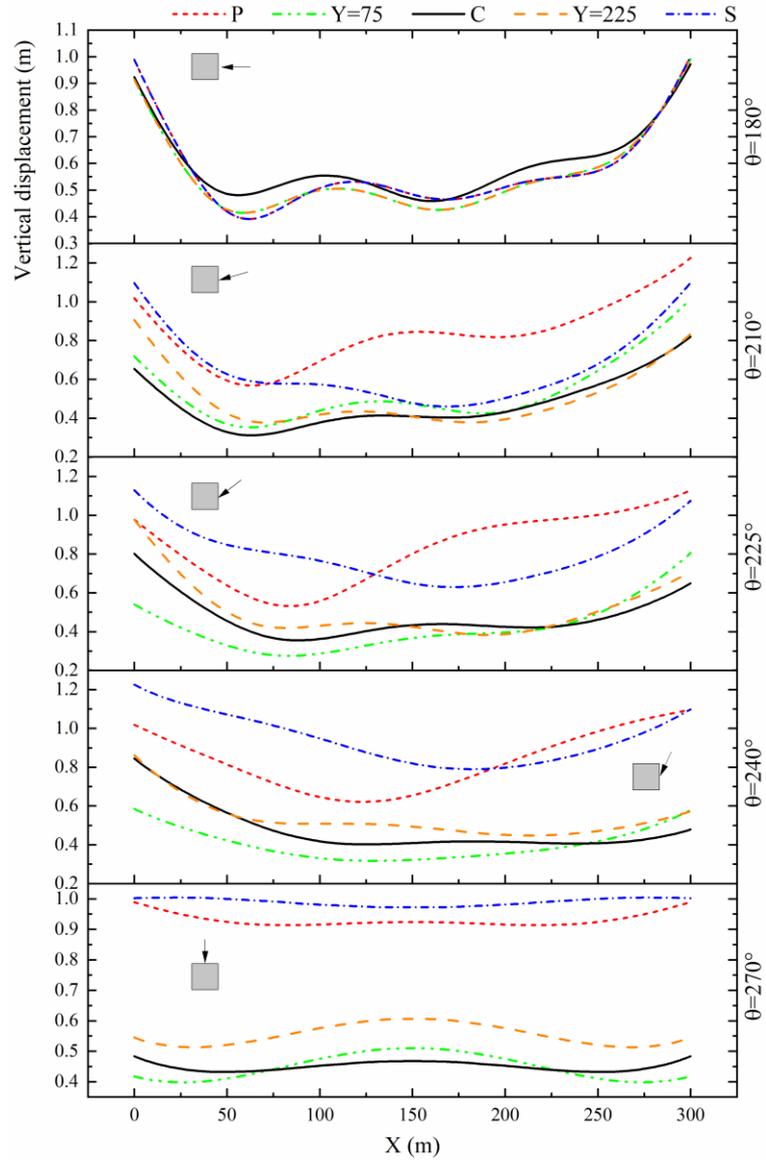

**Fig. D1.** Vertical displacement distribution on P ($Y = 0$), $Y = 75$, C ($Y = 150$), $Y = 225$, S ($Y = 300$) of SV-300 under a 180m regular wave for five incidence angles labeled on the right-side of each sub-figure. The grey square in each subfigure represents SV-300 and the arrow indicates wave incidence angle.

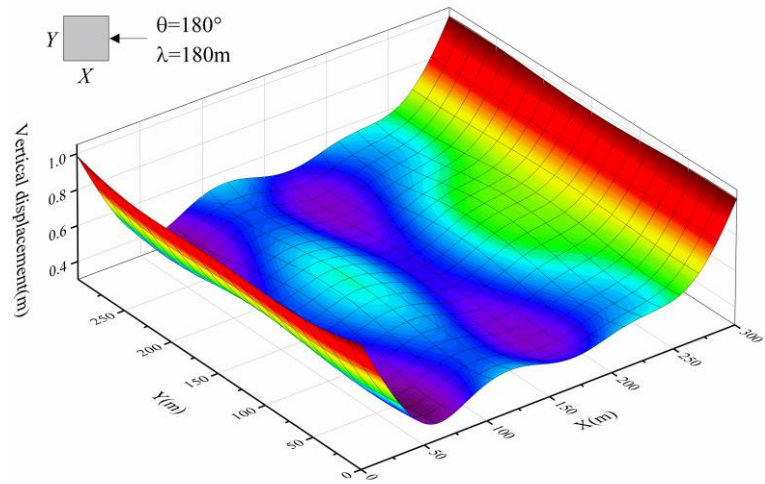

(a) 180m 180degree

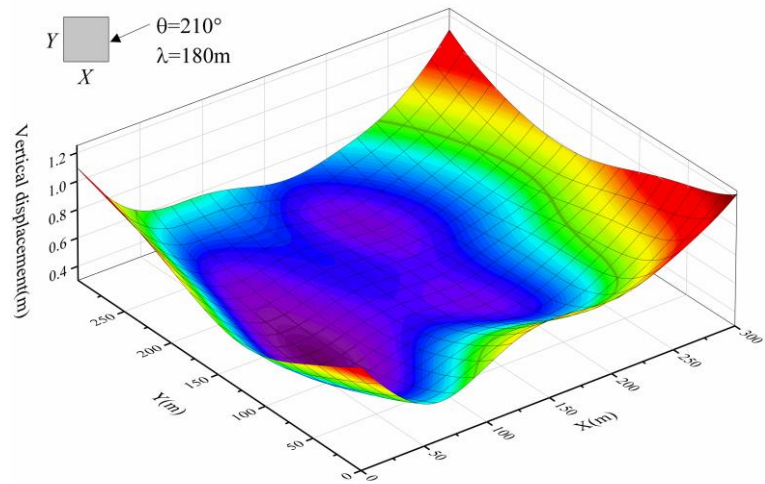

(b) 180m 210degree

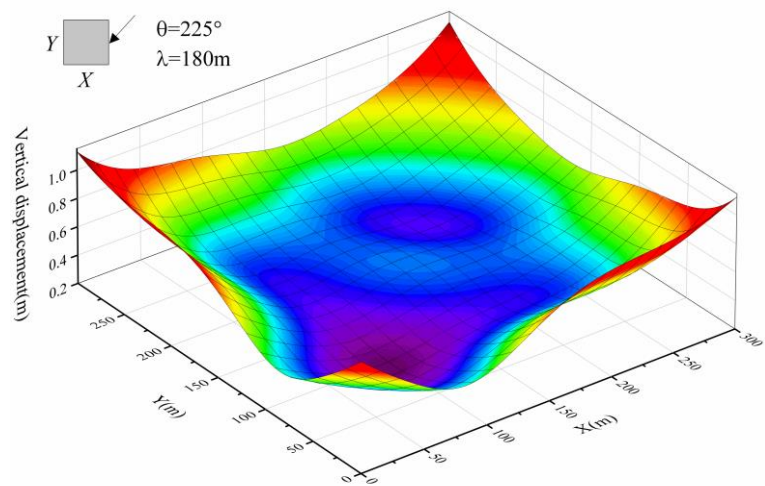

(c) 180m 225degree

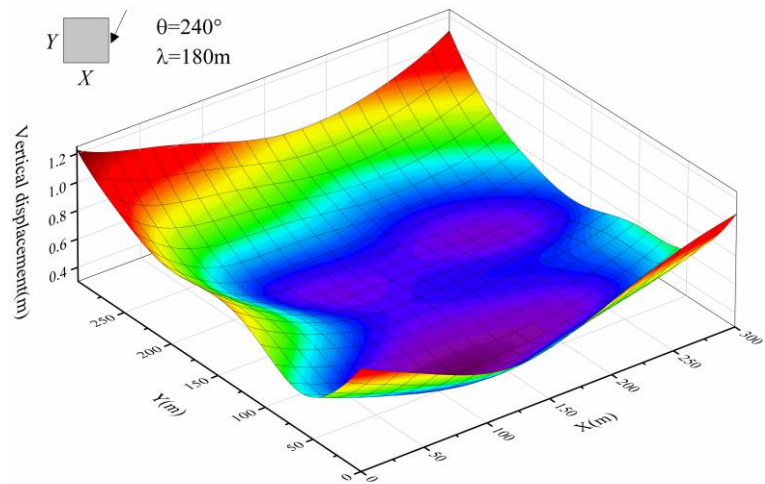

(d) 180m 240degree

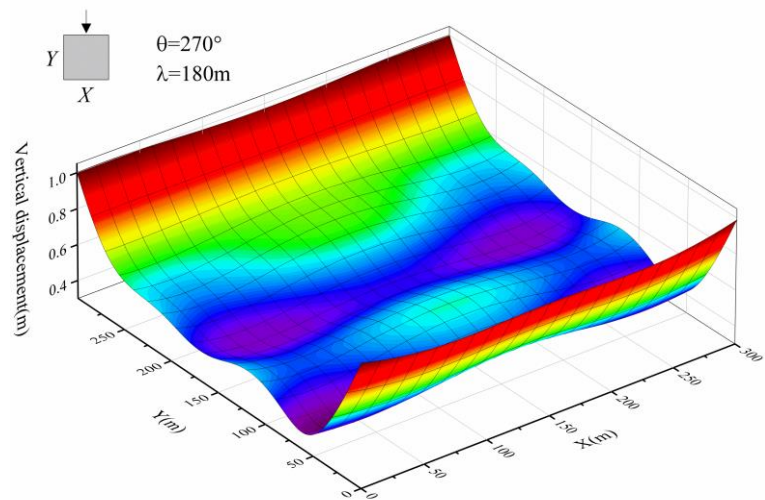

(e) 180m 270degree

**Fig. D2. Vertical displacement distribution along the whole structure under a 180m regular wave for 5 incidence angles. The wave length and wave incidence angle is shown on the left high corner of each subfigure.**

Bending moment results are presented below on SV-300 under regular waves with 225° incidence angle for four wave lengths (120m, 180m, 240m and 300m). Fig. D3 gives bending moment $M_y$ along P, C and S indicated in Fig. 14. Fig. D4 presents bending moment distribution along the whole structure.

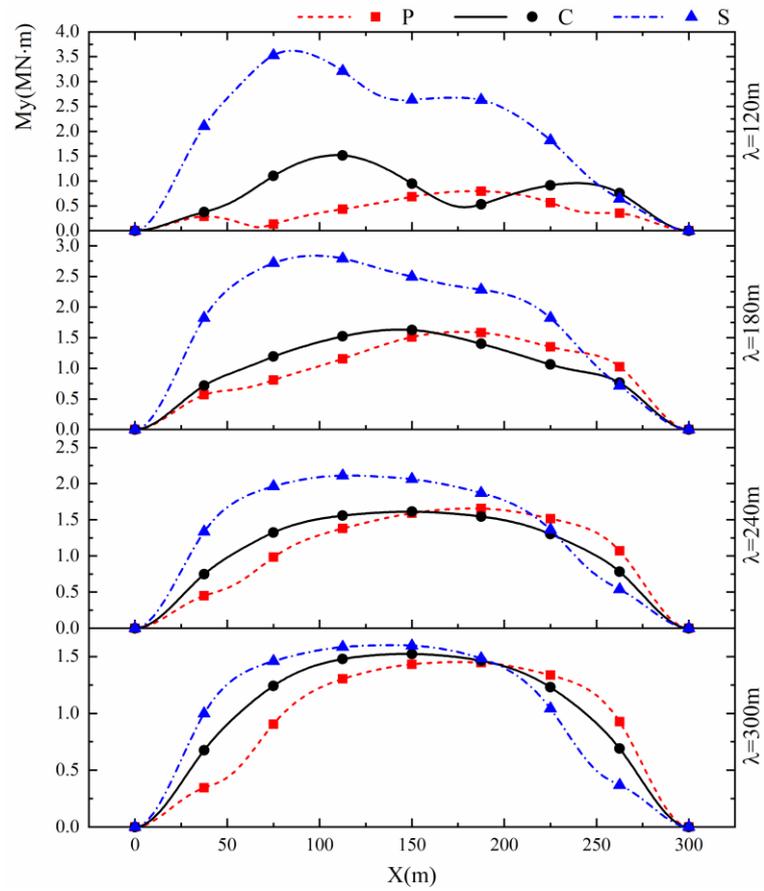

**Fig. D3.** Bending moment distribution $M_y$ on SV-300 under regular waves with a 225-degree incidence angle and four wavelengths (120m, 180m, 24m and 300m), which are shown on the right-side of each subfigure.

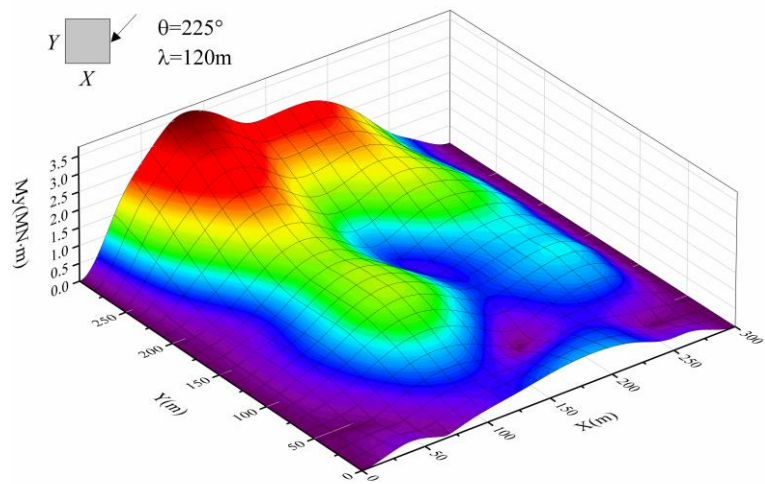

(a) 120m 225degree

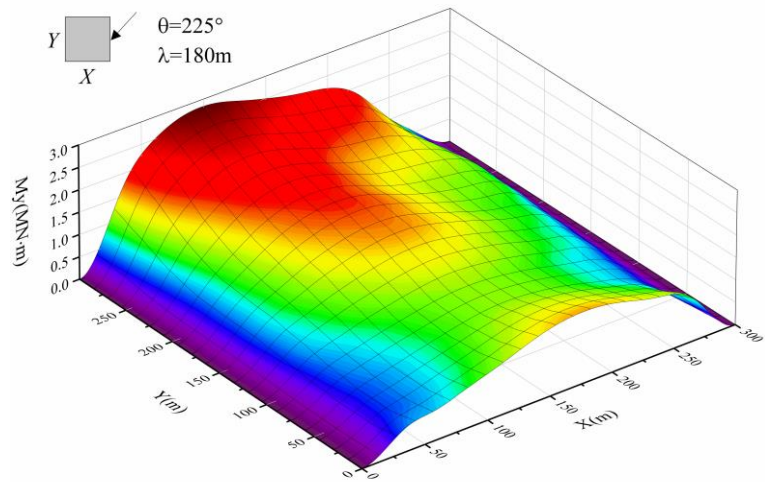

(b) 180m 225degree

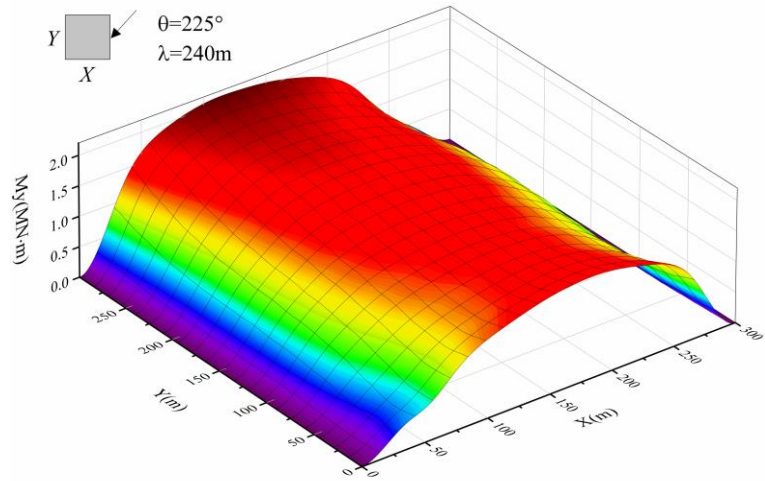

(c) 240m 225degree

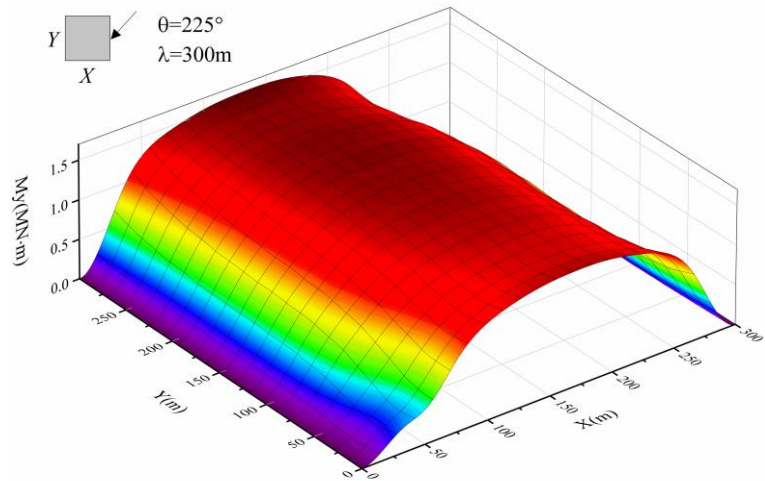

(d) 300m 225degree

**Fig. D4. Bending moment distribution along SV-300 for four wave lengths, which are shown on the left high**

# Appendix E

Say a total of $p$ nodes enter the calculation, which means an amount of $p$ shear force values $N_{qy}^{(m,n)}(x_i, y_i), q = 1, \cdots, p$ and bending moment values $M_{qy}^{(m,n)}(x_q, y_q), q = 1, \cdots, p$ are known. $(x_q, y_q)$ is the coordinate of Node $q$ in the body-fixed coordinate system. Construct $L_M$ and $L_N$ as

$$\begin{cases} L_M(a_1, \ldots, a_{10}) = \sum_{q=1}^{p} \left( \sum_{j=1}^{10} a_j \varphi_j(x_q, y_q) - M_{qy}^{(m,n)}(x_q, y_q) \right)^2 \\ L_N(a_1, \ldots, a_{10}) = \sum_{q=1}^{p} \left( \sum_{j=1}^{10} a_j \frac{\partial}{\partial x} \varphi_j(x_q, y_q) - N_{qy}^{(m,n)}(x_q, y_q) \right)^2 \end{cases} \quad (E1)$$

Apparently, the sum of $L_M$ and $L_N$ should satisfy

$$\frac{\partial}{\partial a_j}(L_M + L_N) = 0, \quad j = 1, 2, \ldots, 10 \quad (E2)$$

Eq. E2 not only ensures coincidence between the distribution surface $M_y^{(m,n)}(x, y)$ and bending moment values at specific boundary nodes, but regulate its trend at these locations with its partial derivative to $x$ equals to the shear force values.

It can be calculated that

$$\begin{aligned} \frac{1}{2}\frac{\partial}{\partial a_k} L_M &= \sum_{q=1}^{p} \left( \sum_{j=1}^{10} a_j \varphi_j(x_q, y_q) - M_{qy}(x_q, y_q) \right) \varphi_k(x_q, y_q) \\ &= \sum_{j=1}^{10} a_j \sum_{q=1}^{p} \varphi_j(x_q, y_q)\varphi_k(x_q, y_q) - \sum_{q=1}^{p} M_{qy}(x_q, y_q)\varphi_k(x_q, y_q) \end{aligned} \quad (E3)$$

$$\frac{1}{2}\frac{\partial}{\partial a_k}L_N = \sum_{q=1}^{p}\left(\sum_{j=1}^{10}a_j\frac{\partial}{\partial x}\varphi_j(x_q,y_q) - N_{qy}(x_q,y_q)\right)\frac{\partial}{\partial x}\varphi_k(x_q,y_q)$$

$$= \sum_{j=1}^{10}a_j\sum_{q=1}^{p}\frac{\partial}{\partial x}\varphi_j(x_q,y_q)\frac{\partial}{\partial x}\varphi_k(x_q,y_q) - \sum_{q=1}^{p}N_{qy}(x_q,y_q)\frac{\partial}{\partial x}\varphi_k(x_q,y_q)$$

(E4)

It is defined for brevity that

$$\begin{cases}(\varphi_k,\varphi_j) = \left(\varphi_j(x_q,y_q),\varphi_k(x_q,y_q)\right) = \sum_{q=1}^{p}\varphi_j(x_q,y_q)\varphi_k(x_q,y_q) \\ (\varphi_k,M_q) = \left(M_{qy}(x_q,y_q),\varphi_k(x_q,y_q)\right) = \sum_{q=1}^{p}M_q(x_q,y_q)\varphi_k(x_q,y_q) \\ (\partial\varphi_k,\partial\varphi_j) = \left(\frac{\partial}{\partial x}\varphi_j(x_q,y_q),\frac{\partial}{\partial x}\varphi_k(x_q,y_q)\right) = \sum_{q=1}^{p}\frac{\partial}{\partial x}\varphi_j(x_q,y_q)\frac{\partial}{\partial x}\varphi_k(x_q,y_q) \\ (\partial\varphi_k,N_q) = \left(N_{qy}(x_q,y_q),\frac{\partial}{\partial x}\varphi_k(x_q,y_q)\right) = \sum_{q=1}^{p}N_q(x_q,y_q)\frac{\partial}{\partial x}\varphi_k(x_q,y_q)\end{cases}$$

(E5)

Substitution of Eq. E5 into Eq. E3 gives

$$\frac{1}{2}\frac{\partial}{\partial a_k}L_M = \sum_{j=1}^{10}a_j(\varphi_k,\varphi_j) - (\varphi_k,M_{qy}) \quad k = 1,2,\cdots 10$$

(E6)

Similarly, with Eq. E5 substituted into Eq. E4

$$\frac{1}{2}\frac{\partial}{\partial a_k}L_N = \sum_{j=1}^{10}a_j(\partial\varphi_k,\partial\varphi_j) - (\partial\varphi_k,N_{qy}) \quad k = 1,2,\cdots 10$$

(E7)

Define $\mathbf{G}_M$, $\mathbf{G}_N$, $\boldsymbol{\varphi}_M$ and $\boldsymbol{\varphi}_N$ as follows:

$$\mathbf{G}_M = \begin{bmatrix} (\varphi_1,\varphi_1) & (\varphi_1,\varphi_2) & (\varphi_1,\varphi_3) & (\varphi_1,\varphi_4) & (\varphi_1,\varphi_5) & (\varphi_1,\varphi_6) & (\varphi_1,\varphi_7) & (\varphi_1,\varphi_8) & (\varphi_1,\varphi_9) & (\varphi_1,\varphi_{10}) \\ (\varphi_2,\varphi_1) & (\varphi_2,\varphi_2) & (\varphi_2,\varphi_3) & (\varphi_2,\varphi_4) & (\varphi_2,\varphi_5) & (\varphi_2,\varphi_6) & (\varphi_2,\varphi_7) & (\varphi_2,\varphi_8) & (\varphi_2,\varphi_9) & (\varphi_2,\varphi_{10}) \\ (\varphi_3,\varphi_1) & (\varphi_3,\varphi_2) & (\varphi_3,\varphi_3) & (\varphi_3,\varphi_4) & (\varphi_3,\varphi_5) & (\varphi_3,\varphi_6) & (\varphi_3,\varphi_7) & (\varphi_3,\varphi_8) & (\varphi_3,\varphi_9) & (\varphi_3,\varphi_{10}) \\ (\varphi_4,\varphi_1) & (\varphi_4,\varphi_2) & (\varphi_4,\varphi_3) & (\varphi_4,\varphi_4) & (\varphi_4,\varphi_5) & (\varphi_4,\varphi_6) & (\varphi_4,\varphi_7) & (\varphi_4,\varphi_8) & (\varphi_4,\varphi_9) & (\varphi_4,\varphi_{10}) \\ (\varphi_5,\varphi_1) & (\varphi_5,\varphi_2) & (\varphi_5,\varphi_3) & (\varphi_5,\varphi_4) & (\varphi_5,\varphi_5) & (\varphi_5,\varphi_6) & (\varphi_5,\varphi_7) & (\varphi_5,\varphi_8) & (\varphi_5,\varphi_9) & (\varphi_5,\varphi_{10}) \\ (\varphi_6,\varphi_1) & (\varphi_6,\varphi_2) & (\varphi_6,\varphi_3) & (\varphi_6,\varphi_4) & (\varphi_6,\varphi_5) & (\varphi_6,\varphi_6) & (\varphi_6,\varphi_7) & (\varphi_6,\varphi_8) & (\varphi_6,\varphi_9) & (\varphi_6,\varphi_{10}) \\ (\varphi_7,\varphi_1) & (\varphi_7,\varphi_2) & (\varphi_7,\varphi_3) & (\varphi_7,\varphi_4) & (\varphi_7,\varphi_5) & (\varphi_7,\varphi_6) & (\varphi_7,\varphi_7) & (\varphi_7,\varphi_8) & (\varphi_7,\varphi_9) & (\varphi_7,\varphi_{10}) \\ (\varphi_8,\varphi_1) & (\varphi_8,\varphi_2) & (\varphi_8,\varphi_3) & (\varphi_8,\varphi_4) & (\varphi_8,\varphi_5) & (\varphi_8,\varphi_6) & (\varphi_8,\varphi_7) & (\varphi_8,\varphi_8) & (\varphi_8,\varphi_9) & (\varphi_8,\varphi_{10}) \\ (\varphi_9,\varphi_1) & (\varphi_9,\varphi_2) & (\varphi_9,\varphi_3) & (\varphi_9,\varphi_4) & (\varphi_9,\varphi_5) & (\varphi_9,\varphi_6) & (\varphi_9,\varphi_7) & (\varphi_9,\varphi_8) & (\varphi_9,\varphi_9) & (\varphi_9,\varphi_{10}) \\ (\varphi_{10},\varphi_1) & (\varphi_{10},\varphi_2) & (\varphi_{10},\varphi_3) & (\varphi_{10},\varphi_4) & (\varphi_{10},\varphi_5) & (\varphi_{10},\varphi_6) & (\varphi_{10},\varphi_7) & (\varphi_{10},\varphi_8) & (\varphi_{10},\varphi_9) & (\varphi_{10},\varphi_{10}) \end{bmatrix} \quad (E8)$$

Notice that $\partial\varphi_1 = \partial\varphi_3 = \partial\varphi_6 = \partial\varphi_{10} = 0$

$$\mathbf{G}_N = \begin{bmatrix} 0 & 0 & 0 & 0 & 0 & 0 & 0 & 0 & 0 & 0 \\ 0 & (\partial\varphi_2,\partial\varphi_2) & 0 & (\partial\varphi_2,\partial\varphi_4) & (\partial\varphi_2,\partial\varphi_5) & 0 & (\partial\varphi_2,\partial\varphi_7) & (\partial\varphi_2,\partial\varphi_8) & (\partial\varphi_2,\partial\varphi_9) & 0 \\ 0 & 0 & 0 & 0 & 0 & 0 & 0 & 0 & 0 & 0 \\ 0 & (\partial\varphi_4,\partial\varphi_2) & 0 & (\partial\varphi_4,\partial\varphi_4) & (\partial\varphi_4,\partial\varphi_5) & 0 & (\partial\varphi_4,\partial\varphi_7) & (\partial\varphi_4,\partial\varphi_8) & (\partial\varphi_4,\partial\varphi_9) & 0 \\ 0 & (\partial\varphi_5,\partial\varphi_2) & 0 & (\partial\varphi_5,\partial\varphi_4) & (\partial\varphi_5,\partial\varphi_5) & 0 & (\partial\varphi_5,\partial\varphi_7) & (\partial\varphi_5,\partial\varphi_8) & (\partial\varphi_5,\partial\varphi_9) & 0 \\ 0 & 0 & 0 & 0 & 0 & 0 & 0 & 0 & 0 & 0 \\ 0 & (\partial\varphi_7,\partial\varphi_2) & 0 & (\partial\varphi_7,\partial\varphi_4) & (\partial\varphi_7,\partial\varphi_5) & 0 & (\partial\varphi_7,\partial\varphi_7) & (\partial\varphi_7,\partial\varphi_8) & (\partial\varphi_7,\partial\varphi_9) & 0 \\ 0 & (\partial\varphi_8,\partial\varphi_2) & 0 & (\partial\varphi_8,\partial\varphi_4) & (\partial\varphi_8,\partial\varphi_5) & 0 & (\partial\varphi_8,\partial\varphi_7) & (\partial\varphi_8,\partial\varphi_8) & (\partial\varphi_8,\partial\varphi_9) & 0 \\ 0 & (\partial\varphi_9,\partial\varphi_2) & 0 & (\partial\varphi_9,\partial\varphi_4) & (\partial\varphi_9,\partial\varphi_5) & 0 & (\partial\varphi_9,\partial\varphi_7) & (\partial\varphi_9,\partial\varphi_8) & (\partial\varphi_9,\partial\varphi_9) & 0 \\ 0 & 0 & 0 & 0 & 0 & 0 & 0 & 0 & 0 & 0 \end{bmatrix} \quad (E9)$$

$$\boldsymbol{\varphi}_M = \begin{bmatrix} (\varphi_1, M_q) \\ (\varphi_2, M_q) \\ (\varphi_3, M_q) \\ (\varphi_4, M_q) \\ (\varphi_5, M_q) \\ (\varphi_6, M_q) \\ (\varphi_7, M_q) \\ (\varphi_8, M_q) \\ (\varphi_9, M_q) \\ (\varphi_{10}, M_q) \end{bmatrix} \quad (E10)$$

$$\boldsymbol{\varphi}_N = \begin{bmatrix} 0 \\ (\partial\varphi_2, N_q) \\ 0 \\ (\partial\varphi_4, N_q) \\ (\partial\varphi_5, N_q) \\ 0 \\ (\partial\varphi_7, N_q) \\ (\partial\varphi_8, N_q) \\ (\partial\varphi_9, N_q) \\ 0 \end{bmatrix} \quad (E11)$$

The undetermined coefficients are expressed as

$$\mathbf{a} = \begin{bmatrix} a_1 \\ a_2 \\ a_3 \\ a_4 \\ a_5 \\ a_6 \\ a_7 \\ a_8 \\ a_9 \\ a_{10} \end{bmatrix} \tag{E12}$$

Substitutions of Eqs. E8-E12 into Eq. E2 gives

$$[\mathbf{G}_M + \mathbf{G}_N]\mathbf{a} = [\boldsymbol{\varphi}_M + \boldsymbol{\varphi}_N] \tag{E13}$$

Then the coefficients are given by

$$\mathbf{a} = [\mathbf{G}_M + \mathbf{G}_N]^{-1}[\boldsymbol{\varphi}_M + \boldsymbol{\varphi}_N] \tag{E14}$$